\documentclass[12pt, draftclsnofoot, onecolumn]{IEEEtran}
\IEEEoverridecommandlockouts

\makeatletter
\def\markboth#1#2{\def\leftmark{\@IEEEcompsoconly{\sffamily}\MakeUppercase{\protect#1}}%
\def\rightmark{\@IEEEcompsoconly{\sffamily}\MakeUppercase{\protect#2}}}
\makeatother

 \PassOptionsToPackage{bookmarks={false}}{hyperref}
\usepackage[utf8x]{inputenc}
\usepackage[english]{babel}
\usepackage{ucs}
\usepackage{amsmath}
\usepackage{amsfonts}
\usepackage{amssymb}
\usepackage{amsthm}
\usepackage{array}
\usepackage{verbatim}
\usepackage{listings}
\usepackage{psfrag}
\usepackage{stfloats}

\usepackage{algorithm}
\usepackage{algorithmic}
\usepackage{url}
  \usepackage{enumerate}
  \usepackage{cite}
\usepackage[usenames,dvipsnames,svgnames,table]{xcolor}

\ifCLASSINFOpdf
  \usepackage[pdftex]{graphicx}
  \usepackage{subfigure}
  \usepackage[outdir=./]{epstopdf}
  \graphicspath{{./figures/}}
   \DeclareGraphicsExtensions{.eps,.ps,.png}
\else
  \usepackage[dvipdf]{graphicx}
  \usepackage{subfigure}
  \graphicspath{{./figures/}}
   \DeclareGraphicsExtensions{.eps,.ps}
\fi

\usepackage[acronyms,nonumberlist,nopostdot,nomain,nogroupskip]{glossaries}
\newacronym{quic}{QUIC}{Quick UDP Internet Connections}
\newacronym{3gpp}{3GPP}{3rd Generation Partnership Project}
\newacronym{adc}{ADC}{Analog to Digital Converter}
\newacronym{5g}{5G}{5th generation}
\newacronym{aimd}{AIMD}{Additive Increase Multiplicative Decrease}
\newacronym{am}{AM}{Acknowledged Mode}
\newacronym{amc}{AMC}{Adaptive Modulation and Coding}
\newacronym{aqm}{AQM}{Active Queue Management}
\newacronym{awgn}{AGWN}{Additive White Gaussian Noise}
\newacronym{balia}{BALIA}{Balanced Link Adaptation}
\newacronym{bdp}{BDP}{Bandwidth-Delay Product}
\newacronym{bf}{BF}{Beamforming}
\newacronym{hbf}{HBF}{Hybrid Beamforming}
\newacronym{cc}{CC}{Congestion Control}
\newacronym{cdf}{CDF}{Cumulative Distribution Function}
\newacronym{pdf}{PDF}{Probability Density Function}
\newacronym{cn}{CN}{Core Network}
\newacronym{cqi}{CQI}{Channel Quality Information}
\newacronym{cp}{CP}{Control Plane}
\newacronym{csirs}{CSI-RS}{Channel State Information - Reference Signal}
\newacronym{dc}{DC}{Dual Connectivity}
\newacronym{dce}{DCE}{Direct Code Execution}
\newacronym{dci}{DCI}{Downlink Control Information}
\newacronym{dl}{DL}{Downlink}
\newacronym{dmr}{DMR}{Deadline Miss Ratio}
\newacronym{dmrs}{DMRS}{DeModulation Reference Signal}
\newacronym{e2e}{E2E}{End-to-End}
\newacronym{ecn}{ECN}{Explicit Congestion Notification}
\newacronym{edf}{EDF}{Earliest Deadline First}
\newacronym{enb}{eNB}{evolved Node Base}
\newacronym{epc}{EPC}{Evolved Packet Core}
\newacronym{es}{ES}{Edge Server}
\newacronym{fdma}{FDMA}{Frequency Division Multiple Access}
\newacronym{fdd}{FDD}{Frequency Division Duplexing}
\newacronym[firstplural=Radio Access Technologies (RATs)]{rat}{RAT}{Radio Access Technology}
\newacronym{fs}{FS}{Fast Switching}
\newacronym{ftp}{FTP}{File Transfer Protocol}
\newacronym{gnb}{gNB}{Next Generation Node Base}
\newacronym{harq}{HARQ}{Hybrid Automatic Repeat reQuest}
\newacronym{hetnet}{HetNet}{Heterogeneous Network}
\newacronym{hh}{HH}{Hard Handover}
\newacronym{hol}{HOL}{Head-of-Line}
\newacronym{ia}{IA}{Initial Access}
\newacronym{imt}{IMT}{International Mobile Telecommunication}
\newacronym{iot}{IoT}{Internet of Things}
\newacronym{los}{LOS}{Line of Sight}
\newacronym{lte}{LTE}{Long Term Evolution}
\newacronym{m2m}{M2M}{Machine to Machine}
\newacronym{mac}{MAC}{Medium Access Control}
\newacronym{mc}{MC}{Multi-Connectivity}
\newacronym{mcs}{MCS}{Modulation and Coding Scheme}
\newacronym{mec}{MEC}{Mobile Edge Cloud}
\newacronym{mi}{MI}{Mutual Information}
\newacronym{mimo}{MIMO}{Multiple-Input Multiple-Output}
\newacronym{mumimo}{MU-MIMO}{Multi-User Multiple-Input Multiple-Output}
\newacronym{mmwave}{mmWave}{millimeter wave}
\newacronym{mr}{MR}{Maximum Rate}
\newacronym{mss}{MSS}{Maximum Segment Size}
\newacronym{mtd}{MTD}{Machine-Type Device}
\newacronym{mtu}{MTU}{Maximum Transmission Unit}
\newacronym{nfv}{NFV}{Network Function Virtualization}
\newacronym{nlos}{NLOS}{Non Line of Sight}
\newacronym{nr}{NR}{New Radio}
\newacronym{ofdm}{OFDM}{Orthogonal Frequency Division Multiplexing}
\newacronym{ofdma}{OFDMA}{Orthogonal Frequency Division Multiple Access}
\newacronym{pdcch}{PDCCH}{Physical Downlink Control Channel}
\newacronym{pucch}{PUCCH}{Physical Uplink Control Channel}
\newacronym{pdcp}{PDCP}{Packet Data Convergence Protocol}
\newacronym{pdsch}{PDSCH}{Physical Downlink Shared Channel}
\newacronym{pdu}{PDU}{Packet Data Unit}
\newacronym{pf}{PF}{Proportional Fair}
\newacronym{pgw}{PGW}{Packet Gateway}
\newacronym{phy}{PHY}{Physical}
\newacronym{pbch}{PBCH}{Physical Broadcast Channel}
\newacronym[plural=\gls{mme}s,firstplural=Mobility Management Entities (MMEs)]{mme}{MME}{Mobility Management Entity}
\newacronym{prb}{PRB}{Physical Resource Block}
\newacronym{pss}{PSS}{Primary Synchronization Signal}
\newacronym{pusch}{PUSCH}{Physical Uplink Shared Channel}
\newacronym{rach}{RACH}{Random Access Channel}
\newacronym{ran}{RAN}{Radio Access Network}
\newacronym{red}{RED}{Random Early Detection}
\newacronym{rf}{RF}{Radio Frequency}
\newacronym{rlc}{RLC}{Radio Link Control}
\newacronym{rlf}{RLF}{Radio Link Failure}
\newacronym{rrc}{RRC}{Radio Resource Control}
\newacronym{rrm}{RRM}{Radio Resource Management}
\newacronym{rr}{RR}{Round Robin}
\newacronym{rs}{RS}{Remote Server}
\newacronym{rsrp}{RSRP}{Reference Signal Received Power}
\newacronym{rss}{RSS}{Received Signal Strength}
\newacronym{rtt}{RTT}{Round Trip Time}
\newacronym{rw}{RW}{Receive Window}
\newacronym{rx}{RX}{Receiver}
\newacronym{sa}{SA}{standalone}
\newacronym{sack}{SACK}{Selective Acknowledgment}
\newacronym{sap}{SAP}{Service Access Point}
\newacronym{sch}{SCH}{Secondary Cell Handover}
\newacronym{scoot}{SCOOT}{Split Cycle Offset Optimization Technique}
\newacronym{sdma}{SDMA}{Spatial Division Multiple Access}
\newacronym{sinr}{SINR}{Signal to Interference plus Noise Ratio}
\newacronym{sm}{SM}{Saturation Mode}
\newacronym{snr}{SNR}{Signal to Noise Ratio}
\newacronym{son}{SON}{Self-Organizing Network}
\newacronym{ss}{SS}{Synchronization Signal}
\newacronym{srs}{SRS}{Sounding Reference Signal}
\newacronym{sss}{SSS}{Secondary Synchronization Signal}
\newacronym{tb}{TB}{Transport Block}
\newacronym{tcp}{TCP}{Transmission Control Protocol}
\newacronym{tdd}{TDD}{Time Division Duplexing}
\newacronym{tdma}{TDMA}{Time Division Multiple Access}
\newacronym{tfl}{TfL}{Transport for London}
\newacronym{tm}{TM}{Transparent Mode}
\newacronym{trp}{TRP}{Transmitter Receiver Pair}
\newacronym{tti}{TTI}{Transmission Time Interval}
\newacronym{ttt}{TTT}{Time-to-Trigger}
\newacronym{tx}{TX}{Transmitter}
\newacronym{ue}{UE}{User Equipment}
\newacronym{ul}{UL}{Uplink}
\newacronym{uml}{UML}{Unified Modeling Language}
\newacronym{um}{UM}{Unacknowledged Mode}
\newacronym{utc}{UTC}{Urban Traffic Control}
\newacronym{vm}{VM}{Virtual Machine}
\newacronym{rsrq}{RSRQ}{Reference Signal Received Quality}
\newacronym{rssi}{RSSI}{Received Signal Strength Indicator}
\newacronym{crs}{CRS}{Cell Reference Signal}
\newacronym{comp}{CoMP}{Coordinated Multi-Point}
\newacronym{cran}{C-RAN}{Cloud \acrlong{ran}}
\newacronym{ca}{CA}{Carrier Aggregation}
\newacronym{cco}{CC}{Carrier Component}
\newacronym{nsa}{NSA}{Non Stand Alone}
\newacronym{embb}{eMBB}{Enhanced Mobility Broadband}
\newacronym{bsr}{BSR}{Buffer Status Report}
\newacronym{srb}{SRB}{Service Radio Bearer}
\newacronym{scm}{SCM}{Spatial Channel Model}
\newacronym{sctp}{SCTP}{Stream Control Transmission Protocol}
\newacronym{mptcp}{MPTCP}{Multi-path TCP}
\newacronym{ietf}{IETF}{Internet Engineering Task Force}
\newacronym{os}{OS}{Operating System}
\newacronym{tls}{TLS}{Transport Layer Security}
\newacronym{rfc}{RFC}{Request for Comments}
\newacronym{http}{HTTP}{HyperText Transfer Protocol}
\newacronym{nat}{NAT}{Network Address Translation}
\newacronym{api}{API}{Application Programming Interface}
\newacronym{rto}{RTO}{Retransmission Timeout}
\newacronym{psc}{PSC}{Public Safety Communication}
\newacronym{rpgm}{RPGM}{Reference Point Group Mobility}
\newacronym{ic}{IC}{Incident Command}
\newacronym{rsu}{RSU}{Road Side Unit}
\newacronym{uav}{UAV}{Unmanned Aerial Vehicle}
\newacronym{iab}{IAB}{Integrated Access and Backhaul}
\newacronym{psd}{PSD}{Power Spectral Density}
\newacronym{mpc}{MPC}{Multi Path Component}
\newacronym{rt}{RT}{Ray Tracer}
\newacronym{aoa}{AoA}{Angle of Arrival}
\newacronym{aod}{AoD}{Angle of Departure}
\newacronym{inr}{INR}{Interference to Noise Ratio}
\newacronym{qd}{QD}{Quasi Deterministic}
\newacronym{wlan}{WLAN}{Wireless Local Area Network}
\newacronym{cad}{CAD}{Computer-aided Design}
\newacronym{ap}{AP}{Access Point}
\newacronym{sta}{STA}{Station}
\newacronym{nrmse}{NRMSE}{Normalized Root Mean Square Error}
\newacronym{ut}{UT}{User Terminal}
\newacronym{bs}{BS}{Base Station}
\newacronym{mmse}{MMSE}{Minimum Mean Squared Error}
\newacronym{gbf}{GBF}{Geometric BeamForming}
\newacronym{cbf}{CBF}{Codebook BeamForming}
\newacronym{fmbf}{FMBF}{Frequency-Flat MMSE BeamForming}
\newacronym{smbf}{SMBF}{Frequency-Selective MMSE BeamForming}
\newacronym{bler}{BLER}{Block Error Rate}
\newacronym{fft}{FFT}{Fast Fourier Transform}
\newacronym{nack}{NACK}{Negative Acknowledgment}
\newacronym{upa}{UPA}{Uniform Planar Array}
\newacronym{tmrs}{TMRS}{TDMA mmWave RR Scheduler}
\newacronym{pmrs}{PMRS}{Padded mmWave RR Scheduler}
\newacronym{amrs}{AMRS}{Asynchronous mmWave almost-RR Scheduler}
\newacronym{rb}{RB}{Resource Block}
\newacronym{udp}{UDP}{User Datagram Protocol}
\newacronym{noma}{NOMA}{Non Orthogonal Multiple Access}
\newacronym{dft}{DFT}{Discrete Fourier Transform}
\newacronym{cav}{CAV}{Connected Autonomous Vehicles}
\newacronym{fr2}{FR2}{Frequency Range 2}
\usepackage{tablefootnote}
\usepackage{booktabs}

\usepackage{tabularx}

\usepackage{tikz}
\usepackage{pgfplots}
\pgfplotsset{compat=newest}
\pgfplotsset{plot coordinates/math parser=false}
\usetikzlibrary{plotmarks,patterns,decorations.pathreplacing,backgrounds,calc,arrows,arrows.meta,spy,matrix}
\usepgfplotslibrary{patchplots,groupplots}
\usepackage{tikzscale}
\usepackage{hyperref}
\usepackage{tikz}
\usepackage{pgfplots}
\pgfplotsset{compat=newest} 
\pgfplotsset{plot coordinates/math parser=false} 
\newlength\fheight
\newlength\fwidth
\usetikzlibrary{plotmarks,patterns,decorations.pathreplacing,backgrounds,calc,arrows,arrows.meta,spy,matrix}
\usepgfplotslibrary{patchplots,groupplots}
\usepackage{tikzscale}

\tikzstyle{startstop} = [rectangle, rounded corners, minimum width=2cm, minimum height=0.5cm,text centered, draw=black]
\tikzstyle{io} = [trapezium, trapezium left angle=70, trapezium right angle=110, minimum width=3cm, minimum height=1cm, text centered, draw=black]
\tikzstyle{process} = [rectangle, minimum width=2cm, minimum height=0.5cm, text centered, draw=black, align=center]
\tikzstyle{decision} = [ellipse, minimum width=2cm, minimum height=1cm, text centered, draw=black]
\tikzstyle{arrow} = [thick,<->,>=stealth]
\tikzstyle{line} = [thick,>=stealth]
\tikzstyle{darrow} = [thick,<->,>=stealth,dashed]
\tikzstyle{sarrow} = [thick,->,>=stealth]
\tikzstyle{larrow} = [line width=0.1mm,dashdotted,<->,>=stealth]

\pgfplotsset{every tick label/.append style={font=\scriptsize}, 
             every axis/.append style={
             width=\fwidth, height=\fheight, at={(0\fwidth,0\fheight)}, 
             xlabel style={font=\footnotesize\color{white!15!black}},
             xmajorgrids,
             ylabel style={yshift=-0.15cm, font=\footnotesize\color{white!15!black}},
             ymajorgrids,
             legend style={font=\footnotesize\color{white!15!black}},
             /pgfplots/ybar legend/.style={/pgfplots/legend image code/.code={\draw[##1,/tikz/.cd,yshift=-0.25em](0cm,0cm) rectangle (10pt,1em);},},
             }}

\definecolor{SchoolColor}{RGB}{0.71, 0, 0.106}
\definecolor{chaptergrey}{rgb}{0.61, 0, 0.09} 
\definecolor{midgrey}{rgb}{0.4, 0.4, 0.4}
\definecolor{chaptergreen}{rgb}{0.09, 0.612, 0}
\definecolor{chapterpurple}{rgb}{0.522, 0, 0.612}
\definecolor{chapterlightgreen}{rgb}{0, 0.612, 0.522}

\makeatletter
\def\grd@save@target#1{%
  \def\grd@target{#1}}
\def\grd@save@start#1{%
  \def\grd@start{#1}}
\tikzset{
  grid with coordinates/.style={
    to path={%
      \pgfextra{%
        \edef\grd@@target{(\tikztotarget)}%
        \tikz@scan@one@point\grd@save@target\grd@@target\relax
        \edef\grd@@start{(\tikztostart)}%
        \tikz@scan@one@point\grd@save@start\grd@@start\relax
        \draw[minor help lines] (\tikztostart) grid (\tikztotarget);
        \draw[major help lines] (\tikztostart) grid (\tikztotarget);
        \grd@start
        \pgfmathsetmacro{\grd@xa}{\the\pgf@x/1cm}
        \pgfmathsetmacro{\grd@ya}{\the\pgf@y/1cm}
        \grd@target
        \pgfmathsetmacro{\grd@xb}{\the\pgf@x/1cm}
        \pgfmathsetmacro{\grd@yb}{\the\pgf@y/1cm}
        \pgfmathsetmacro{\grd@xc}{\grd@xa + \pgfkeysvalueof{/tikz/grid with coordinates/major step x}}
        \pgfmathsetmacro{\grd@yc}{\grd@ya + \pgfkeysvalueof{/tikz/grid with coordinates/major step y}}
        \foreach \x in {\grd@xa,\grd@xc,...,\grd@xb}
        \node[anchor=north] at (\x,\grd@ya) {\pgfmathprintnumber{\x}};
        \foreach \y in {\grd@ya,\grd@yc,...,\grd@yb}
        \node[anchor=east] at (\grd@xa,\y) {\pgfmathprintnumber{\y}};
      }
    }
  },
  minor help lines/.style={
    help lines,
    gray,
    line cap =round,
    xstep=\pgfkeysvalueof{/tikz/grid with coordinates/minor step x},
    ystep=\pgfkeysvalueof{/tikz/grid with coordinates/minor step y}
  },
  major help lines/.style={
    help lines,
    line cap =round,
    line width=\pgfkeysvalueof{/tikz/grid with coordinates/major line width},
    xstep=\pgfkeysvalueof{/tikz/grid with coordinates/major step x},
    ystep=\pgfkeysvalueof{/tikz/grid with coordinates/major step y}
  },
  grid with coordinates/.cd,
  minor step x/.initial=.5,
  minor step y/.initial=.2,
  major step x/.initial=1,
  major step y/.initial=1,
  major line width/.initial=1pt,
}
\makeatother

\newcommand{\Hb}{\mathbf{H}}

\newcommand{\V}{\mathbf{V}}

\newcommand{\I}{\mathbf{I}}

\newcommand{\x}{\mathbf{x}}

\newcommand{\vv}{\mathbf{v}}

\newcommand{\y}{\mathbf{y}}

\newcommand{\w}{\mathbf{w}}

\newcommand{\ab}{\mathbf{a}}

\usepackage{pifont}
\theoremstyle{plain}

   \definecolor{blueH3}{rgb}{0,.5,1}
   \definecolor{blueH2}{rgb}{0,0.25,0.75}
   \definecolor{blueH1}{rgb}{0,0,0.5}   
   \definecolor{grayOldText}{rgb}{.5,.5,.5}
   \definecolor{VCobalt}{HTML}{005682}
   \definecolor{TZTeal}{HTML}{008080}
   \definecolor{KYJade}{HTML}{008151}
   \definecolor{ARust}{HTML}{a10000}
   \definecolor{FFucsia}{HTML}{7000c3}
   
\newcommand{\CASE}[1]{\STATE \textbf{case} #1\textbf{:} \begin{ALC@g}}
\newcommand{\ENDCASE}{\end{ALC@g}}

\newcommand{\DEFAULT}{\STATE \textbf{default:} \begin{ALC@g}}
\newcommand{\ENDDEFAULT}{\end{ALC@g}}
\newcommand{\DEFAULTLINE}[1]{\STATE \textbf{default:} }

\definecolor{Tangerine}{HTML}{d45500}
\glsdisablehyper
\begin{document}

\title{Hybrid Beamforming in 5G mmWave Networks: a Full-stack Perspective}

  \author{
  \IEEEauthorblockN{Felipe G\'omez-Cuba,~\IEEEmembership{Member, IEEE}, Tommaso Zugno,~\IEEEmembership{Student Member, IEEE}, \\Junseok Kim,~\IEEEmembership{Member, IEEE}, Michele Polese,~\IEEEmembership{Member, IEEE}, \\Saewoong Bahk,~\IEEEmembership{Senior member, IEEE}, Michele Zorzi,~\IEEEmembership{Fellow, IEEE}}
   
\thanks{Felipe G\'omez-Cuba is with AtlantTIC, University of Vigo, Spain. Email: gomezcuba@gts.uvigo.es.}
\thanks{Tommaso Zugno and Michele Zorzi are with the Department of Information Engineering, University of Padova, Padova, Italy. Email: \{zugnotom, zorzi\}@dei.unipd.it}
\thanks{Junseok Kim is with System LSI, Samsung Electronics, Gyeonggi-do, South Korea. Email: junseok.kim@samsung.com.}
\thanks{Michele Polese is with the Institute for the Wireless Internet of Things, Northeastern University, Boston, MA USA. Email: m.polese@northeastern.edu.}
\thanks{Saewoong Bahk is with the Department of ECE and INMC, Seoul National University, Seoul, South Korea. Email: sbahk@snu.ac.kr.}

}

\maketitle

\begin{abstract}
This paper studies the cross-layer challenges and performance of \gls{hbf} and \gls{mumimo} in 5G \gls{mmwave} cellular networks with full-stack TCP/IP traffic and MAC scheduling. While previous research on \gls{hbf} and \gls{mumimo} has focused on link-level analysis of full-buffer transmissions, this work reveals the interplay between \gls{hbf} techniques and the higher layers of the protocol stack. 
To this aim, prior work on full stack simulation of \gls{mmwave} cellular network has been extended by including the modeling of \gls{mumimo} and \gls{hbf}. Our results reveal novel relations between the networking layers and the \gls{hbf} \gls{mumimo} performance in the physical layer. Particularly, throughput can be increased in 5G networks by means of \gls{sdma}. However, in order to achieve such benefits it is necessary to take into account certain trade-offs and the implementation complexity of a full-stack \gls{hbf} solution.
\end{abstract}


\glsresetall
\glsunset{nr}

\section{Introduction}
\label{sec:intro}


The next generations of mobile networks will need to satisfy the capacity, latency, and reliability demands of present and future networked applications~\cite{itu-r-2083}. On one hand, the connected society is heavily relying on wireless networks to consume a wide range of digital services, from video streaming to cloud storage, which require high capacity links. On the other hand, remote control applications for smart factories and eHealth introduce high reliability and low latency constraints. To satisfy the wide range of use cases, the design of flexible and highly performing 5G and beyond cellular networks is of primary importance~\cite{BocHLMP:14,giordani2020toward}.

To this end, the \gls{3gpp} has recently defined a set of specifications for \gls{3gpp} \gls{nr}~\cite{38300}, a new \gls{ran} architecture that includes several technological advancements with respect to previous generations~\cite{BocHLMP:14}. It features a flexible frame structure, in which the parameters of the \gls{ofdm} numerology can be tuned to support different services, and, for the first time, the possibility of using the \gls{mmwave} spectrum in the radio access. Releases 15 and 16, indeed, support carrier frequencies up to 52.6 GHz, with possible extensions to 71 GHz planned for Release 17~\cite{qualcomm201971}. 

The \gls{mmwave} spectrum is a promising technological enabler that can provide high capacity links, thanks to the wide availability of bandwidth, which would help combating the spectrum crunch of traditional sub-6 GHz bands~\cite{RanRapE:14}. \gls{nr} will thus be capable of exploiting much larger bandwidths with respect to previous \gls{3gpp} waveforms, with up to 400 MHz for each carrier~\cite{38300}.
Another feature that \glspl{mmwave} introduce is the possibility of packing a large number of antenna elements in a small form factor, both in base stations and in mobile devices. This makes it possible to focus the energy of the transmission in narrow beams, and to offset the increase in path loss (due to the higher carrier frequency) with a higher \gls{bf} gain~\cite{SunRap:cm14}. Different \gls{bf} architectures have also been considered in the literature. With analog \gls{bf}, the transceivers are equipped with a single \gls{rf} chain, and a single beam is generated using analog phase shifters in the $N$ antenna elements of the phase array. More advanced transceivers use hybrid or digital \gls{bf} architectures, with $K \le N$ \gls{rf} chains. While increasing the complexity and power consumption of the device, they enable a finer control on the \gls{bf} process, which can be based on combined digital and analog processing~\cite{Heath:partialBF}.

Hybrid and digital \gls{bf} architectures, therefore, are capable of steering multiple beams from a single antenna array, with (possibly) independent data streams, effectively enabling \gls{mumimo} operations at \glspl{mmwave}~\cite{SunRap:cm14}. As a result, this makes it possible to increase the spectral efficiency of the network, as different users can be served with \gls{sdma} in the same time and frequency resources.
\gls{hbf} solutions, in particular, are considered as a cost- and energy-effective solution for \gls{mumimo} at \glspl{mmwave}, and have been practically implemented and deployed in commercial devices~\cite{mondal2019reconfigurable}. 

\subsection{Contributions}

The specifications for 3GPP NR natively consider the possibility of \gls{mumimo} transmissions, and embed a beam management framework to support directional transmissions~\cite{8458146}. However, despite the promising features of \gls{hbf} at \glspl{mmwave}, the state of the art currently lacks an analysis of how a \textit{physical layer} based on \gls{hbf} interacts with the \textit{full protocol stack}, from the \gls{mac} layer, e.g., for scheduling, to the transport layer and applications.

To fill this gap, in this paper we present an analysis of how \gls{hbf} techniques can be integrated in the protocol stack of 5G and beyond cellular networks, focusing on optimal beam design and scheduling strategies for different scenarios and applications. To the best of our knowledge, this is the first contribution that performs an end-to-end evaluation of the potential and challenges of \gls{mumimo} \gls{hbf} with a full-stack perspective. Notably, we
\begin{itemize}
	\item provide a tutorial on how the protocol stack in \gls{3gpp} \gls{nr} supports \gls{mumimo}, in terms of basic signaling and operations, with references to the relevant technical specifications;
	\item propose two \gls{mmse} \gls{hbf} design strategies for \gls{mumimo} in \gls{3gpp} \gls{nr}, comparing approaches which present a trade-off between complexity and performance, and how they can be practically integrated in the 5G protocol stacks;
	\item unravel the sub-optimal interaction that may arise from the combination of sophisticated \gls{mumimo} \gls{hbf} strategies and traditional \gls{mac} scheduling strategies, which do not fully support \gls{sdma} over independent beams. We then propose scheduling strategies which, while being standard compliant, improve the performance of \gls{mumimo} communications;
	\item present the results of an extensive simulation campaign, based on our novel implementation of \gls{hbf} \gls{mumimo} transmissions in the ns-3 \gls{mmwave} module~\cite{mezzavilla2018end}. To the best of our knowledge, this tool\footnote{Available at \url{https://github.com/signetlabdei/ns3-mmwave-hbf}} is the first open source software to support the simulation of \gls{hbf} \gls{mumimo} at \glspl{mmwave}, with a 3GPP-like protocol stack, the \gls{3gpp} channel model for \gls{mmwave} frequencies, and, thanks to the integration with ns-3, the TCP/IP stack and realistic applications. Our results show that \gls{hbf} can improve the performance of a saturated, single-layer mmWave network, for a variety of end-to-end traffic flows, and that an \gls{hbf}-aware scheduling design is fundamental to achieve the potential gains of multi-layer solutions. 
\end{itemize}

\subsection{State of the Art}
\label{sec:soa}

\gls{mumimo} channels have received considerable academic interest since the 90s \cite{goldsmith2002fundamental}. This interest increased even further as the number of antennas per device grew, harnessing the advantages of \textit{massive \gls{mimo}} \cite{4176578}. Nowadays \gls{mumimo} with very large antenna arrays is an integral part of the 5G \gls{nr} cellular standard published by the \gls{3gpp} \cite{TS38211v16,TS38212v16,TS38213v16}. \gls{mumimo} pilots, channel estimation and feedback procedures in the standard are concisely described in \cite{8692922}. Signal processing techniques for \gls{mmwave} systems are reviewed in \cite{7400949}. Beam-management procedures in 5G are covered in \cite{8458146}. Moreover, the research community has widely studied efficient transceiver architectures~\cite{sohrabi2017hybrid}, and beam design strategies~\cite{Heath:partialBF}, and has characterized the performance gains that \gls{mumimo} and \gls{sdma} can yield in \gls{mmwave} deployments~\cite{kulkarni2016comparison}.

The research community has also recently focused on the evaluation of the end-to-end performance of \gls{5g} mmWave networks, with analysis~\cite{moltchanov2020analytical}, simulations~\cite{choi20195g}, and experiments~\cite{Sur:2017:WGW:3117811.3117817}. This has been recognized as fundamental to enable a proper design and validation of 5G technologies, given that new, unexplored technologies are being coupled with protocols and networks that have not been specifically designed for NR~\cite{mezzavilla2018end}. For example, the authors of~\cite{8613277} investigate how \gls{tcp} behaves on top of mmWave links, highlighting pitfalls and strategies to improve the overall performance. The state of the art, however, has mostly focused on the interaction between the higher layer of the protocol stack and a physical layer with analog \gls{bf}~\cite{slezak2018understanding}. Some evaluations are available for \gls{hbf} at \glspl{mmwave}, but focus on simplified full-buffer link-layer evaluations~\cite{han2015large,sohrabi2017hybrid,Heath:partialBF,kulkarni2016comparison}. Similarly, performance analysis of digital \gls{bf} schemes for 3GPP NR has focused on power consumption and control procedures~\cite{dutta2020case,8458146}. 

As discussed in the previous paragraphs, this paper advances the state of the art by jointly analyzing \gls{hbf} design and higher layer issues, to understand how their interplay can be optimized to maximize the performance of the network. 

\subsection{Paper Structure}

The remainder of the paper is organized as follows. Sec.~\ref{sec:system} discusses \gls{hbf} and scheduling design in \gls{3gpp} \gls{nr} networks. Sec.~\ref{sec:perf_eval}, then, describes the performance evaluation results, and Sec.~\ref{sec:conclusions} concludes the paper, providing suggestions for future research directions.

\section{Full-stack Integration of \gls{hbf} for \gls{mmwave}s}
\label{sec:system}

This section describes how the NR protocol stack is designed to support \gls{hbf} (Sec.~\ref{sec:nr-hbf}), reviews \gls{hbf} design strategies (Sec.~\ref{sec:bfdesign}), and analyzes the challenges associated with the interplay between \gls{hbf} and schedulers, proposing different \gls{hbf}-aware scheduling policies (Sec.~\ref{sec:schdesign}).

\subsection{\gls{hbf} in the \gls{3gpp} \gls{nr} Stack}
\label{sec:nr-hbf}

The \gls{3gpp} 5G standard specifies the \gls{nr} waveform and conformance requirements for devices \cite{TS38211v16,TS38102v16}. The interface between the waveform and the antenna array is standardized through a series of ``antenna ports." The details of \gls{bf} operations applied to signals in each port are left to the vendor implementation, and constrained only by conformance requirements such as those in \cite{TS38102v16}.

In the \gls{nr} waveform, complex symbols are mapped in a 3-dimensional \gls{ofdm} resource grid comprising the \gls{ofdm} symbol number in time ($n$), the \gls{ofdm} subcarrier number in frequency ($k$), and the \textit{\gls{sdma} layer} number ($\ell$) \cite[Table 7.3.1.3-1]{TS38211v16}. Furthermore, layers can be mapped to more than one antenna port ($p$) using several precoding configurations \cite{TS38211v16}. These antenna ports are defined as signal input/output interfaces to the antenna array. The mapping of antenna ports to actual antenna elements may be vendor-specific, but must guarantee that in the same port and within the same ``slot'' of 14 consecutive \gls{ofdm} symbols, the channel may be inferred through \glspl{dmrs} that are orthogonal and specific in each port \cite{TS38212v16}. We note that the standard does not mandate that each port corresponds to a frequency-flat analog beam, in other words, further vendor-specific frequency-selective \gls{hbf} operations can be considered.

At the \gls{nr} \gls{mac} layer, the scheduler assigns radio \glspl{rb} in the grid with indices $(n,k,\ell)$. The broadest time division are \textit{frames} of $10$ ms duration. Each frame is divided in 10 \textit{subframes} of $1$ ms. These large scale time units are similar to \gls{lte}, and can admit either \gls{tdd} or \gls{fdd} configurations. However, differently from \gls{lte}, slots in any subframe can be labeled as \gls{dl}, \gls{ul} or \textit{flexible}, where the latter represent an innovation in 5G that permits the scheduler to dynamically change the \gls{dl}/\gls{ul} division over consecutive subframes.

In the smaller time-scale of the \gls{ofdm} signal, the dimensions of the time-frequency grid depend on the fundamental \textit{numerology} parameter $\mu\in\{0,1,2,3,4\}$. \gls{ofdm} symbols are grouped in \textit{slots} of 14 symbols, such that each subframe has $2^\mu$ slots. The inter-carrier spacing is $\Delta f = 2^\mu \times 15$ kHz, and each \gls{ofdm} symbol's duration is $\frac{2^{-\mu}}{14}$ ms. The maximum bandwidth without carrier aggregation is 400~MHz. In addition, the smallest scheduling granularity in \gls{nr} is the ``mini-slot," which can be only 2 \gls{ofdm} symbols long and does not necessarily have to be time-aligned with multiples of the nominal slot start instants. This allows \gls{nr} schedulers to assign ``asynchronous" transmissions that do not start at the same time. Moreover, parallel resource grids are defined simultaneously for each port $p$ \textbf{and for each numerology $\mu$}, meaning that it is possible to schedule contiguous transmissions with different physical layer parameters in the same frame.

The NR versatile waveform supports different frequencies, from the conventional $700$ MHz--$6$ GHz spectrum up to the $24-70$ GHz \gls{mmwave} spectrum. As a result, some \gls{nr} options are not useful when operating at \glspl{mmwave}. One is \gls{ofdma}, since highly directive \gls{mmwave} \gls{bf} typically requires frequency-flat analog operations. This means that the radio device cannot apply different analog beams to different subcarriers of the same \gls{ofdm} symbol. Therefore, in \gls{mmwave} all subcarriers $k$ in a port-symbol pair need to be assigned to the same user. As a consequence, scheduling in \gls{mmwave} \gls{nr} reduces to a 2-dimension \gls{tdma} and \gls{sdma} grid $(n,\ell)$. The second is the use of \gls{sdma} transmissions with more than one layer to the same user, since typical \gls{mmwave} \gls{mimo} channel matrices are rank deficient (i.e., the second largest eigenvalue is much smaller than the first) \cite{SunRap:cm14}. 
In turn, multiple transmissions with rank $\geq 2$ to the same user are ineffective, thus \gls{sdma} can only be implemented as a \gls{mumimo} technique, but not as a single user \gls{mimo} technique.

\subsection{\gls{hbf} Design}
\label{sec:bfdesign}

We assume that simultaneous \gls{sdma} transmissions are allocated into different layers $\ell$. The channel matrix between the \gls{bs} and each \gls{ue} $u$ is denoted by $\Hb_{u}[n,k]$ in \gls{ofdm} symbol $n$ and subcarrier $k$. In \gls{dl}, the BS selects a wideband analog \gls{bf} vector for each transmit layer $\vv_\ell$ using some \gls{bf} scheme, and the UE receives with the analog \gls{bf} vector $\w_u$. Thus the \textit{effective} scalar complex channel between the transmit layer $\ell$ and the receive antenna port of the UE is given by
$$h_{eq}[u,\ell,n,k]=\w_u^T\Hb_{u}[n,k]\vv_\ell,$$
and the \gls{ul} channel is computed with the transposed channel matrix and swapping transmitter and receiver beamforming vectors, resulting in the same complex scalar number.

We follow a \gls{sinr}-based point-to-point link performance model. This means that we obtain an expression for the \gls{sinr} of each link, assume that simultaneous links are decoded separately, and map the \gls{sinr} of each link to the \gls{bler} of the associated transmission. This is a simplification of real decoding hardware that makes the simulation of a large network tractable. Real \gls{nr} demodulation and decoding may use sophisticated joint decoding such as \gls{mumimo} sphere decoding \cite{fincke1985improved}, combined with the LDPC and Polar channel codes of each transmission \cite{TS38211v16}.

For multiple simultaneous downlink transmissions sent by the \gls{bs} on multiple layers,
we write the \gls{dl} \gls{sinr} of user $u$ at subcarrier $k$ as a function of the effective channel gains as
\begin{equation}
\label{eq:sinrdl}
    SINR_u^{DL}[k]=\frac{ L_u|h_{eq}[u,\ell(u),n,k]|^2P/K}{\sum_{u'\neq u} L_{u}|h_{eq}[u,\ell(u'),n,k]|^2P/K+\Delta f N_o}
\end{equation}
where $K$ is the number of subcarriers, $L_u$ is the pathloss of user $u$, $P/K$ is the BS transmitted power per layer equally divided among all subcarriers, $N_o$ is the noise \gls{psd} and $\Delta f$ is the inter-carrier spacing. For the discussion of \gls{bf} design at a single BS, we evaluate the sum of interference only over \glspl{ue} connected to the same BS, producing self-interference. If there are other BSs in the scenario, their \gls{bf} design is independent and their interference may be written as a constant in addition to $N_o$. We note that the self-interference terms correspond to ``mismatched'' beams, i.e., an interfering signal is received by user $u$ through the channel of user $u$ but it was sent by the transmitter layer $\ell(u')$ using a \gls{bf} vector designed for user $u'$. Usually this would mean that an interference signal's power is lower than the desired signal power. Still, this interference term may be significantly stronger than the noise, and thus the link \gls{sinr} will be much lower than the \gls{snr}.

In \gls{ul} self-interference is even more severe because the \gls{ul} \gls{sinr} expression is in turn
\begin{equation}
\label{eq:sinrul}
SINR_u^{UL}[k]=\frac{ L_u|h_{eq}[u,\ell(u),n,k]|^2P}{\sum_{u'\neq u} L_{u'}|h_{eq}[u',\ell(u),n,k]|^2P+\Delta f N_o},
\end{equation}
where the layer assigned to user $u$ receives the interference signal of layer $\ell(u')$ through the channel of user $u'$. In other words, pathloss gains in the denominator $L_{u'}$ are also mismatched. Since the pathloss gain of an interferer might occasionally be much greater than the pathloss gain of the desired signal $L_{u}$, \gls{bf} alone is not enough to guarantee that the interference power is lower than the desired signal power in \gls{ul} as it was in \gls{dl}. For example this would happen in a layer that is receiving from a user far from the \gls{bs} that is coexisting with a nearby user in another layer.

Under our \gls{sinr} point-to-point link model, we propose the following design methods to obtain \gls{bf} vectors to achieve good link \gls{snr} or \gls{sinr} values in \eqref{eq:sinrdl} and \eqref{eq:sinrul}. Our \gls{bf} designs extend beyond the standard precoding tables \cite{TS38211v16}, but they can be implemented within the vendor-specific part of the \gls{bf} system, where ports are mapped to antenna elements:

\subsubsection{\gls{gbf}}

We denote the antenna array response as a function $\ab(\theta,\phi)$ that depends on the angles of azimuth and elevation $(\theta,\phi)$. For example, in a \gls{upa} with $N_1 \times N_2$ antennas separated half a wavelength, the $i$-th coefficient of this vector would be $$a_i(\theta,\phi)=e^{-j\frac{\pi}{2}\left((i \mod N_1)\sin(\theta) + \lfloor i / N_1\rfloor\sin(\phi)\right)},$$
whereas other expressions would give $\ab(\theta,\phi)$ for other antenna array shapes.

Notably, in the \gls{upa} and other array shapes where $\ab(\theta,\phi)^H\ab(0,0)=\ab(0,0)^H\ab(-\theta,-\phi)$, we can easily adopt the vector $\ab(\theta,\phi)^H$ to  design a beam that points in the direction $(\theta,\phi)$. In geometric \gls{bf}, the vectors are simply selected by pointing the array in the physical direction between the BS position and the UE position. That is, if the position coordinates are $(x_{BS},y_{BS},z_{BS})$ and $(x_u,y_u,z_u)$, then we have
    \begin{equation}
    \begin{split}
    \theta_D&=\arctan \frac{y_u-y_{BS}}{x_u-x_{BS}} + \pi \mathrm{I}_{(x_u<x_{BS})} \mod 2\pi \\
    \phi_D&=\arctan \frac{z_u-z_{BS}}{\sqrt{(y_u-y_{BS})^2+(x_u-x_{BS})^2}} \\
    \theta_A&=\theta_D+\pi\\
    \phi_A&=-\phi_D\\
    \vv_{\ell(u)}^T&=\ab(\theta_D,\phi_D)^H\\
    \w_u^T&=\ab(\theta_A,\phi_A)^H\\
    \end{split}
    \end{equation}
    where $\mathrm{I}_{(e)}$ is the indicator function for event $e$, the subindex $D$ indicates the angles of departure, subindex $A$ indicates the angles of arrival, and the subindex $\ell(u)$ indicates the layer assigned to the transmission to UE $u$.
    
    Finally, \gls{gbf} vectors are analog, so layers are matched one-to-one with array ports in this scheme. A potential shortcoming of this \gls{bf} model is that it does not adapt to changes in the channel matrix. The strongest channel gain associated with $\Hb_{u}[n,k]$ is represented by its largest singular value and associated eigenvectors, which could be significantly different from the physical geometric direction between the devices, especially in NLOS channels.  
    
\subsubsection{\gls{cbf}}


We denote a \textit{\gls{bf} codebook} $\mathcal{B}$ as a small collection of possible \gls{bf} vectors (either because of array hardware limitations or because the feedback is limited to a log$_2|\mathcal{B}|$ bit message). The transmitter sends reference signals using all the vectors in $\mathcal{B}_D$, and the receiver tests decoding the reference signals with all vectors in $\mathcal{B}_A$. Finally, the receiver chooses a pair of vectors from each codebook based on the observations. For example, the receiver may choose the vectors that result in the maximum empirically-observed reference signal power. In our implementation we assume the following ideal max-\gls{snr} criterion
\begin{equation}
\vv_{\ell(u)},\w_u=\arg \max_{ \vv \in\mathcal{B}_D,\w \in\mathcal{B}_A} |\w_u^T\Hb_{u}[n,k_{ref}]\vv_\ell|^2,
\end{equation}
where $k_{ref}$ is a single subcarrier index where we assume a narrowband reference signal is carried. Finally, the receiver would only need to send to the transmitter a beam indicator message describing the index that $\vv_{\ell(u)}$ occupies in the look-up table containing $\mathcal{B}_D$. 

Codebook \gls{bf} vectors are analog and layers are matched one-to-one with array ports, as in \gls{gbf}. Thanks to the use of a simple codebook-lookup technique, feedback overhead would be very low. 
A potential drawback is that by using a single-subcarrier reference $|\w_u^T\Hb_{u}[n,k_{ref}]\vv_\ell|^2$ the selection does not take into account the gains that would be experienced by any other subcarrier  $|\w_u^T\Hb_{u}[n,k]\vv_\ell|^2 \forall k\neq k_{ref}$. This means that only the \gls{snr} of the reference is maximized while that of other subcarriers is not. Nonetheless, due to the fact that the \gls{mmwave} channel matrix is rank-deficient and the beams in the codebook are rather coarse, the \gls{snr} in different subcarriers can be quite similar and this shortcoming is not too severe.

\subsubsection{\gls{fmbf}}

\gls{gbf} and \gls{cbf} focus solely on maximizing the effective channel gains between the BS and UE $u$ through its associated layer $\ell(u)$. This, in turn, maximizes user $u$'s link \gls{snr}, but does not account for interference to and from other \glspl{ue}, which could result in low \gls{sinr} even if the \gls{snr} is high.

 Therefore, to improve the \gls{sinr}, we introduce a low-complexity, low-dimensional linear matrix mapping between layers and ports, in combination with an auxiliary analog \gls{cbf} underlying scheme. 
 Let us denote the \gls{bf} vectors selected using \gls{cbf} by $\w_u^{CB}$ and $\vv_\ell^{CB}$, and by $h_{eq}[u,p,n,k_{ref}]=(\w_u)^T\Hb_{u}[n,k_{ref}]\vv_p$ the complex channel coefficients observed between user $u$ and port $p$ at the reference subcarrier $k_{ref}$.

We assume that first the system conducts a codebook exploration as in \gls{cbf} and loads the best codebook \gls{bf} vector for each user $u$ to different antenna ports denoted $p(u)$. In addition, we assume that right after the codebook exploration the \gls{bs} notifies each user of all the vectors of interest, and the receivers estimate the effective complex scalars $\sqrt{L_u}h_{eq}^{CB}[u,p(u'),n,k_{ref}]$ for all pairs $(u,p(u)')$. To report these auxiliary effective channel coefficients back to the BS, since a single reference subcarrier is used, would require $N_u^2 N_{bit}$ bits of feedback, where $N_{bit}$ is the number of bits used to encode each complex number and $N_u$ is the number of simultaneous users. For example in a scenario with $N_u=4$ the feedback would be 1024 bits with high precision 32-bit floating point encoding, or 96 bits with a more aggressive 3-bit quantizer.

To simplify the notation, we assume in this section that the active users are numbered sequentially $u\in\{0\dots N_u\}$ and that their assigned layer and port numbers are equally sequential $\ell(u)=p(u)=u$. We also omit the \gls{ofdm} symbol index $n$. Using the auxiliary scalar channel coefficients the BS builds the following \gls{mumimo} reference equivalent channel matrix:
\begin{equation}
\label{eq:equivchanmmse}
\Hb_{eq}[k_{ref}]=\left(\begin{array}{cccc}
\sqrt{L_1}h_{eq}^{CB}[1,1,k_{ref}] & \sqrt{L_1}h_{eq}^{CB}[1,2,k_{ref}] & \dots & \sqrt{L_1}h_{eq}^{CB}[1,N_p,k_{ref}]\\
\sqrt{L_2}h_{eq}^{CB}[2,1,k_{ref}] & \sqrt{L_2}h_{eq}^{CB}[2,2,k_{ref}] & \dots & \sqrt{L_2}h_{eq}^{CB}[2,N_p,k_{ref}]\\
\vdots & \vdots & \ddots & \vdots\\
\sqrt{L_{N_u}}h_{eq}^{CB}[N_u,1,k_{ref}] & \sqrt{L_{N_u}}h_{eq}^{CB}[N_u,2,k_{ref}] & \dots & \sqrt{L_{N_u}}h_{eq}^{CB}[N_u,N_p,k_{ref}]\\
\end{array}\right)    
\end{equation}
where $N_p=N_u$ is the number of analog \gls{bf} ports, each associated to a single user. Moreover since $\ell(u)=p(u)=u$, the desired channels are in the main diagonal of this matrix. Finally, on the receiver side, the receiving \gls{bf} vectors would remain those of \gls{cbf}, while on the transmitter side the BS designs a precoding matrix matching layers to ports by building the following \gls{mmse} \gls{dl} precoding matrix
$$\V_{\gls{mmse}}[k_{ref}]=\Hb_{eq}^H(\Hb_{eq}\Hb_{eq}^H+\frac{N_o\Delta f}{P}\I)^{-1}.$$
We adopt the \gls{mmse} technique because, when the noise is weak compared to the transmitted power, then $\frac{N_o\Delta fK}{P}\to 0$, and the expression converges to the pseudoinverse (zero-forcing precoder), i.e., $\Hb_{eq}[k_{ref}]\V_{\gls{mmse}}[k_{ref}]=\I$, thus suppressing the interference. In addition when the noise is strong, in the limit $\frac{N_o\Delta fK}{P}\to \infty$, the \gls{mmse} expression converges to the hermitian (matched filter) which maximizes the received SNR. Thus, \gls{mmse} offers a balance between interference suppression and noise reduction giving good \gls{sinr} values for any noise-to-transmitted power ratio ($\frac{N_o\Delta fK}{P}$). 


Finally, the final \textit{effective} transmit \gls{bf} vectors at the \gls{bs} for \gls{dl} are obtained by first computing
$$\left(\tilde\vv_1^{\gls{mmse}}\dots \tilde\vv_{N_u}^{\gls{mmse}}\right)= \left(\vv_1^{CB}\dots \vv_{N_u}^{CB}\right)\V_{\gls{mmse}}[k_{ref}],$$
and then introducing the following normalization to preserve the transmitted power constraint in each layer:
$$\vv_u^{\gls{mmse}}=\tilde\vv_u^{\gls{mmse}}/|\tilde\vv_u^{\gls{mmse}}|.$$
Introducing these effective vectors into \eqref{eq:sinrdl}, instead of the auxiliary \gls{cbf} vectors we discussed earlier, results in the new \gls{sinr} values of the \gls{mmse} technique.

For \gls{ul}, an equivalent hybrid combining at the BS receiver can be formulated by adopting the transpose of the matrices described above. 

The \gls{fmbf} technique relies solely on estimations performed at the reference subcarrier $k_{ref}$ and is still frequency-flat. This introduces only a small amount of additional feedback as only one subcarrier equivalent matrix needs to be reported. The precoding/combining matrix is explicitly designed to improve the \gls{sinr} in the reference subcarrier and, as a side effect, the \gls{sinr} would improve in other subcarriers with similar equivalent channel matrices. But since \gls{fmbf} does not take into account the effective channel of the other subcarriers, it does not guarantee complete interference suppression in all subcarriers.

\subsubsection{\gls{smbf}}

\gls{smbf} is an \gls{hbf} scheme comprising a frequency-selective low-dimensional linear matrix mapping from ports to layers, which may be different in each subcarrier, together with a codebook based frequency-flat analog beam selection to map ports to antenna array elements. 

For this, in \gls{dl} we need to assume that after beam codebook exploration is performed, pilot signals are transmitted in all subcarriers and the receivers can report  back to the transmitter a large set of effective channel coefficients $\{h_{eq}^{CB}[u,p(u'),n,k]$ for all pairs $(u,p(u'))$ and subcarrier indices $k\}$. This would require roughly $K\times N_u^2N_{bit}$ bits of feedback. For example if $K=100$, for the case $N_u=4$ and with a coarse quantizer with $N_{bit}=3$, we could send the resulting $9.6$ kbits of feedback in a single OFDM symbol, whereas using high precision complex number encoding with $N_{bit}=32$ would be prohibitive. Using the effective channel information the transmitter builds a collection of $K$ different equivalent channel matrices, one for each subcarrier ($\Hb_{eq}[k]\forall k\in\{1\dots K\}$). For each subcarrier $k$ the transmitter designs a different digital precoding matrix $$\V_{\gls{mmse}}[k]=\Hb_{eq}^H[k](\Hb_{eq}[k]\Hb_{eq}[k]^H+\frac{N_o\Delta f}{P}\I)^{-1}.$$
Thus in the \gls{smbf} scheme the precoding matrix that maps antenna ports to layers is different in each subcarrier. Finally, normalization and calculation of effective \gls{bf} vectors proceeds as in the \gls{fmbf} case, but with the effective vectors introduced in \eqref{eq:sinrdl} taking different values for each subcarrier index $k$. For \gls{ul} the same considerations regarding the transpose channel matrix at the receiver are applied.

This frequency-selective approach requires the estimation of effective channel coefficients for all pairs of co-existing users and for all subcarriers, which represents $K\times$ more feedback than in the \gls{fmbf} scheme. In return, the frequency-selective approach guarantees an explicit suppression of the inter-user interference in all subcarriers. Potentially, this means that the \gls{sinr} can be almost as high as the SNR without interference of a 1-layer system.

\subsubsection{Future Work} We leave for future extensions of our work the study of \gls{mumimo} techniques that go beyond linear \gls{bf} and independent encoding and decoding in each link. In recent years \gls{noma} \cite{6692652} has gained attention in the literature. \gls{noma} schemes can be regarded as practical implementations of the classic capacity-achieving schemes for the \gls{mac} and broadcast channels in information theory \cite{goldsmith2002fundamental}. As mentioned above, in \gls{ul} the \gls{bs} could decode all incoming transmissions jointly using Sphere Decoding \cite{fincke1985improved} or Successive Interference Cancellation. Conversely, in \gls{dl}, the \gls{bs} could jointly encode all transmitted signals using schemes inspired by Dirty Paper Coding, such as lattice-based Quantization Index Modulation \cite{923725}. Moreover, we also leave for future work the study of \gls{bf} performance degradation due to quantization in the channel estimations versus the amount of incurred feedback overhead.

\subsection{\gls{hbf} and Scheduling Interaction}
\label{sec:schdesign}

We assume that the scheduler allocates transmissions in a 2-dimensional resource grid combining \gls{tdma} and \gls{sdma}. All subcarriers in the same \gls{ofdm} symbol are allocated to the same UE due to the fact that the \gls{bf} system in \gls{mmwave} is at least partially frequency-flat. The scheduler produces allocation decisions periodically for each slot of 14 symbols. The standard supports flexible configurations for allocating control information, i.e., the \gls{pdcch}, in specific regions of each frame \cite{TS38213v16}. We assume a periodical control signaling scheme where, for every 14-symbol slot, the first symbol always contains the \gls{pdcch}. In the \gls{pdcch}, \gls{dci} control messages are delivered to all users.
The symbols 2 to 13 are used for data and marked as ``flexible," meaning that they can be employed for \gls{dl} or \gls{ul} in any slot and this choice may vary over different slots. Finally, in the 14-th symbol in each slot the UEs transmit \gls{ul} control information to the BS.

\begin{figure}
    \centering
    \subfigure[\gls{bf} conflict with different transmission start times: At $T_1$ allocation $\#1$ does not observe reference signals from other layers. The \gls{bs} cannot estimate \eqref{eq:equivchanmmse} and use \gls{mmse} \gls{bf}, falling back to \gls{cbf}. When Allocation $\#2$ starts at $T_2$, the transmission of reference signals by Allocation $\#1$ has already passed, so Allocation $\#2$ must fall back to \gls{cbf} as well. In addition, Allocation $\#1$ cannot change its \gls{bf} configuration at $T_2$ either. After $T_2$, both allocations experience the interference power of a \gls{cbf} scenario even though \gls{mmse} is supported by all devices.]{
        \includegraphics[width=.45\columnwidth]{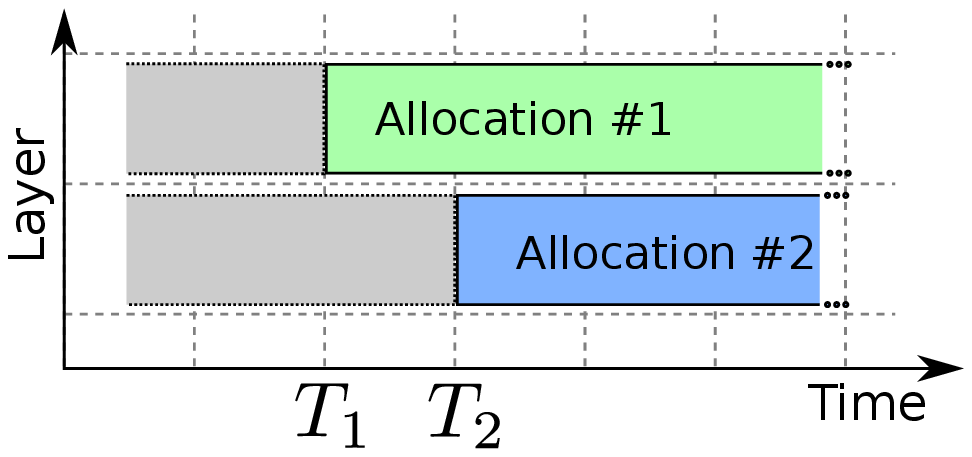}
        \label{fig:bfschedasync}
    }
    \hspace{.1in}
    \subfigure[\gls{bf} conflict with forced simultaneous transmission start: The transmission in the top layer ends at $T_1$, but the scheduler leaves a padding symbol without signal and the new allocation starts at $T_2$. When transmission in the bottom layer ends at $T_2$, both layers start a new allocation simultaneously. Both allocations can observe the other's reference signals and estimate the off-diagonal coefficients of \eqref{eq:equivchanmmse}. \gls{mmse} \gls{bf} can be employed and interference is reduced. However, the frame resource region corresponding to the time interval $T_2-T_1$ in the top layer is wasted. ]{
        \includegraphics[width=.45\columnwidth]{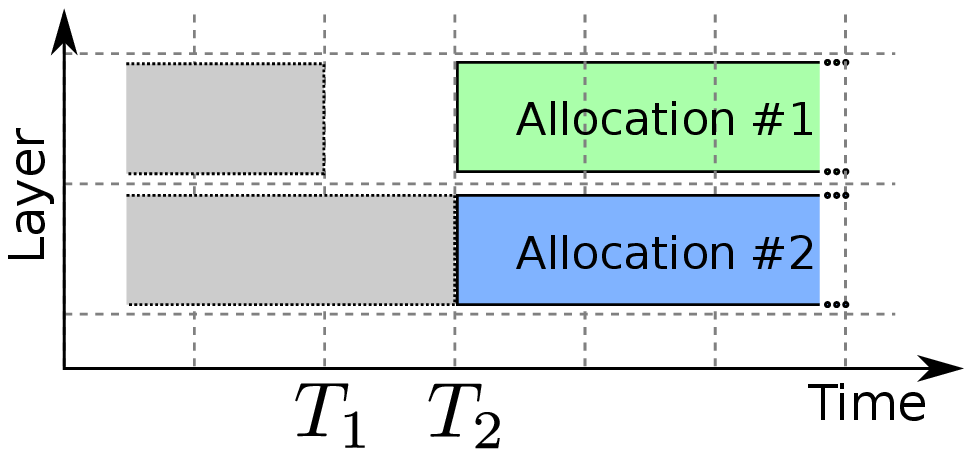}
        \label{fig:bfschedsync}
    }
    \caption{Example of \gls{mmse} \gls{bf} conflict with different transmission start times.}
    \label{fig:bfschedconflict}
    \vspace{-0.4cm}
\end{figure}

\begin{figure}
    \centering
    \subfigure[PMRS: The resources in gray are wasted as padding to guarantee that simulataneous allocations always start at the same time.]{
        \includegraphics[width=.45\columnwidth]{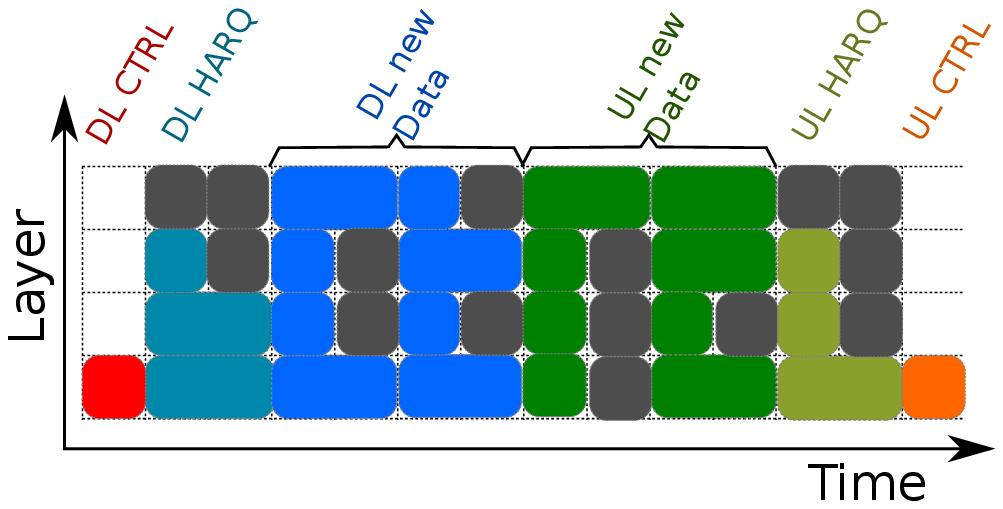}
    \label{fig:padschedexamples}
    }
    \hspace{.1in}
    \subfigure[AMRS: The resources in white are still available in the sense that, if more users were added, these resources could be assigned to them.]{
        \includegraphics[width=.45\columnwidth]{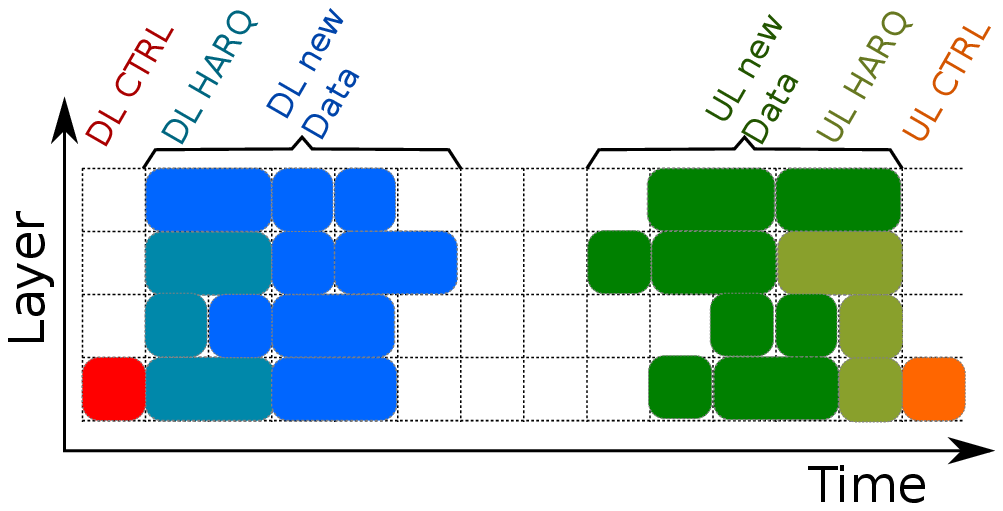}
    \label{fig:asyncschedexamples}
    }
    \caption{Examples of scheduler slot decisions with our two proposals}
    \label{fig:schedexamples}
    \vspace{-0.4cm}
\end{figure}

We assume perfect channel estimation and do not model the \glspl{dmrs} explicitly. Since the smallest scheduling unit is a 2-symbol mini-slot with 1 \gls{dmrs} symbol \cite{TS38211v16}, in our model we assume that the minimum data allocation unit is reduced to 1 symbol of data transmission. We assume that allocated transmissions on different layers may present different start times (in symbol index units). Since each allocated transmission has only one front-loaded \gls{dmrs}, the \gls{bf} configuration of each transmission must be selected at the start of the transmission and cannot vary over the duration of the same transmission. This means that, for a pair of overlapping transmissions that start at different instants, the transmission that started first does not have information on the interference to design its \gls{bf} (Fig. \ref{fig:bfschedasync}). Therefore we assume that \gls{mmse} precoding/combining can only be applied to groups of allocations that start at the same time (Fig. \ref{fig:bfschedsync}). Thus, there is a conflict between scheduling constraints and the applicability of the \gls{mmse} technique, as illustrated in Fig. \ref{fig:bfschedconflict}. We consider two approaches illustrated in Fig. \ref{fig:schedexamples}: the first can make full use of \gls{mmse} but imposes additional constraints on the scheduler leading to lower resource efficiency (Fig. \ref{fig:padschedexamples}). In the second, we allow the scheduler to freely allocate resources even with different start times among different transmissions, achieving a more efficient frame resource occupation  (Fig. \ref{fig:asyncschedexamples}), but causing a fraction of the transmission events to be unable to use \gls{mmse} for interference reduction.

\subsubsection{\gls{pmrs}}

This scheduler guarantees that possibly overlapping transmissions start at the same time in all layers. To do so, given $N_\ell$ layers, $N_s$ symbols and $N_u$ total UEs, the scheduler first divides the subframe equally in $N_b=\lceil N_u/N_\ell \rceil$ ``\gls{sdma} bundles." Each \gls{sdma} bundle is defined as a collection of up to $N_\ell$ concurrent transmissions with the same start time, but allocated to different layers. The bundles are further time-multiplexed using \gls{tdma} over the full subframe, where each bundle is exactly $N_a=\lfloor N_s/N_b \rfloor$ symbols long in time. All layers in the bundle start transmitting at the same time but may end transmitting at different times, according to the amount of data each \gls{ue} has to transmit. Indeed, within each bundle, and for each layer, one UE is selected. If this UE demands fewer than $N_a$ symbols, then its transmission ends before the end of the bundle, and the remaining symbols are left blank (padding). If $N_u>N_s\times N_\ell$, then some UEs are left unserved and become the first UEs in the list for the next subframe in a \gls{rr} fashion. $N_u<N_s\times N_\ell$ then all UEs get equal allocations.

\gls{pmrs} follows a \gls{tdma} first and \gls{sdma} second principle, guaranteeing equal start times for all transmissions in a bundle. This guarantees that \gls{mmse} \gls{bf} is always usable and interference will fully depend on the chosen \gls{bf} scheme. The padding part in each bundle constitutes wasted symbols, and thus this scheduler may display some inefficiency in resource occupation.

\subsubsection{\gls{amrs}}

This scheduler, instead, does not waste any symbol in padding: given $N_\ell$ layers, $N_s$ symbols and $N_u$ total UEs, the scheduler first divides the users into $N_\ell$ ``user groups." Each user group is defined as a collection of UEs served in the same layer. The UEs of the same group are further time-multiplexed using an independent \gls{tdma} RR technique for each layer without taking into account decisions for other layers. If all UEs demand more resources than available, then each UE receives $\lfloor N_s N_\ell/N_u\rfloor$ symbols. However, UEs that demand fewer resources receive fewer symbols, their allocations end sooner, and the next UE in the same group begins its allocation immediately after, without any padding. Due to this, the start times of transmissions in one layer are determined independently of the start times of other layers. Notice that the \glspl{ue} are divided among the layers using an integer division and, as a result, the number of \glspl{ue} per layer may differ by one unit. Finally, \glspl{ue} that cannot get any resources will be served first in the next slot. Since allocations may have different sizes we call the scheduler ``almost-\gls{rr}."

The \gls{amrs} thus follows an \gls{sdma} first and \gls{tdma} second principle, guaranteeing that no symbols are left blank wastefully (i.e., when there is still demand). Free symbols may exist when the total demand of all UEs is lower than $N_sN_\ell$, but these symbols are not ``blocked'' and would be allocated if there were more demand. However, equal start times for all concurrent transmissions are not guaranteed. This means that \gls{mmse} \gls{bf} is not fully used, and interference will be a mixture of two types of events: some allocations that just by chance happen to start at the same time will still use \gls{mmse}, and some others that start at different times will not. Therefore, interference can be significantly greater than the best-case scenario of the selected \gls{bf} scheme.

\section{Performance Evaluation and Tradeoffs}
\label{sec:perf_eval}

This section presents a comprehensive performance evaluation of \gls{hbf} schemes, evaluating the tradeoffs in the design of \gls{bf} and scheduling strategies. We first describe the ns-3-based, end-to-end, full-stack simulator that we developed (Sec.~\ref{sec:ns3}), and the simulation scenario (Sec.~\ref{sec:scenario}), and then present results to compare the different \gls{bf} strategies and scheduling schemes, according to different metrics and configurations of the protocol stack (Sec.~\ref{sec:bfcomparison}-Sec.~\ref{sec:sources}).

\subsection{End-to-end Simulation of \gls{hbf} for \gls{mmwave}}
\label{sec:ns3}

We implemented an \gls{mumimo} \gls{hbf} extension for the ns-3 \gls{mmwave} module introduced in \cite{mezzavilla2018end}. Earlier releases of this module were implemented using the NYU mmwave channel model \cite{7501500}, and adopting the \gls{lte} module in the official ns-3 release as a base \cite{nsnamweb}. Besides the bulk of the multiple-layer implementation, we have introduced adjustments that bring our simulation closer to the \gls{3gpp} 5G \gls{nr} standards. Instead of the NYU channel model, we adopt the most recent \gls{3gpp} channel model implementation in ns-3 \cite{zugno2020implementation}. In addition, the \gls{ofdm} resource grid parameters (bandwidth, subcarrier spacing, symbol duration, and number of slots per frame) reflect those of \gls{nr}, as described in Sec.~\ref{sec:nr-hbf} and~\cite{TS38211v16}. Notice that the ns-3 mmWave module assumes that control signaling is ideal and messages are never lost or corrupted.

In our implementation we have introduced modifications to numerous C++ classes in the ns-3 mmwave module.
Notably, the antenna array module now supports multiple antenna ports, with different \gls{bf} configurations. Moreover, the 3GPP channel model implementation has been extended to account for the multi-layer interference of Eq.~\eqref{eq:sinrdl} and Eq.~\eqref{eq:sinrul}, while the channel abstraction code and the physical layer implementation have been refactored to support multiple \gls{sdma} asynchronous layers (i.e., transmissions from a single entity). The \gls{bf} strategies described in Sec.~\ref{sec:bfdesign} have been implemented in a plug-and-play fashion, leveraging a novel, flexible \gls{bf} module. Finally, we updated the ns-3 mmwave module \gls{mac} layer to support multiple asynchronous layers, by properly accounting for the mapping of upper layer PDUs to mmwave Transport Blocks on different antenna ports, the management of \gls{harq} retransmissions, the CQI estimation, and the control signaling. The \gls{mac} layer also features a plug-and-play implementation of the schedulers introduced in Sec.~\ref{sec:schdesign}, which is backward compatible and allows comparison with the other scheduling strategies implemented in the ns-3 mmWave module~\cite{mezzavilla2018end}. We refer the reader to the publicly available Github repository with the \gls{hbf} extension for additional details.

\subsection{Simulation Scenario}
\label{sec:scenario}

\begin{table}
    \caption{Device Radio Configuration}
    \vspace{-0.4cm}
    \label{tab:radioconf}
    \centering
    \begin{tabular}{lllll}\toprule
           & Transmit Power & Noise Figure & Number of Layers & Antenna Array    \\\midrule 
        BS & $30$ dBm  & $5$ dB       & 1 or 4      & $8\times 8$ UPA  \\
        UE & $30$ dBm  & $5$ dB       & 1           & $4\times 4$ UPA  \\\bottomrule
    \end{tabular}
    \vspace{-0.4cm}
\end{table}

In the next few subsections we present different simulations pertaining to different aspects of the \gls{sdma} \gls{mumimo} \gls{mmwave} system. For all simulations below, we consider a random \gls{mmwave} cellular system with one BS located at the origin of the coordinates (0,0) with a height of $25$~m, and 7 UEs located at random positions uniformly distributed in a disc of radius $100$~m with a height of $1.6$~m. We generate 20 such random deployments and average the results over the random UE locations and channels. We assume that due to the considerable pathloss in \gls{mmwave}, inter-cell interference is severely attenuated and it is sufficient to simulate one cell. This is different from prior work that simulated 4G systems, where the simulators had to consider also a set of ``encircling" neighboring cells to model interference realistically \cite{fgomez2014improvedrelaying}.

We configured the \gls{nr} \gls{ofdm} waveform with numerology $\mu=2$, which corresponds to a subcarrier spacing of $60$~kHz. The system operates at $28$~GHz central frequency with a bandwidth of $198$~MHz divided into 275 \glspl{rb}, each including 12 subcarriers. There are 4 slots per subframe with duration  250 $\mu$s, and the \gls{ofdm} symbol duration is $17.85$ $\mu$s including the CPs. We adopt the channel model described in \gls{3gpp} TR~38.901~\cite{TR38901} and consider the ``Urban Macro'' scenario. The radio hardware in the devices was configured with the parameters listed in Table~\ref{tab:radioconf}.

\subsection{Comparison of \gls{bf} Solutions}
\label{sec:bfcomparison}

\begin{figure}[t]
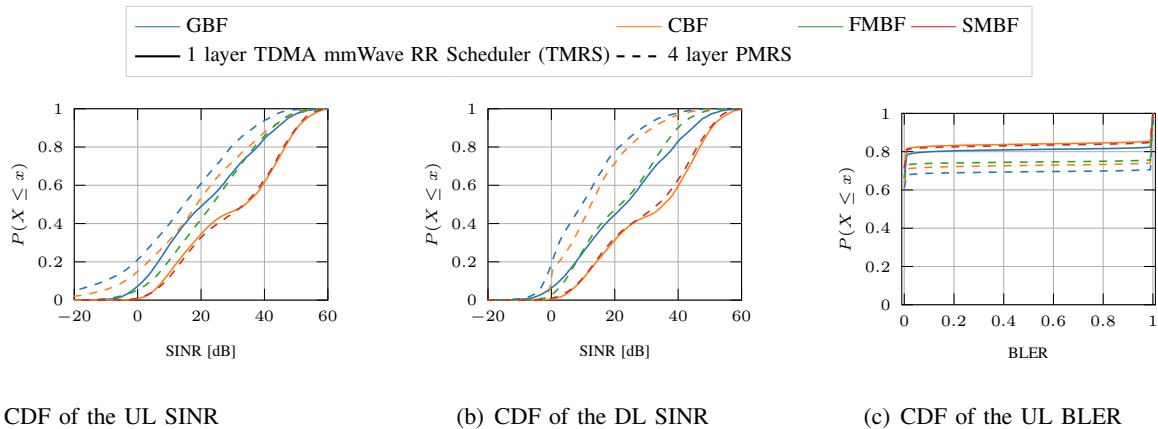

    \centering
      \setlength\fwidth{.3\columnwidth}
      \setlength\fheight{0.1\columnwidth}
\begin{tikzpicture}

\definecolor{color1}{rgb}{1,0.498039215686275,0.0549019607843137}
\definecolor{color2}{rgb}{0.172549019607843,0.627450980392157,0.172549019607843}
\definecolor{color4}{rgb}{0.580392156862745,0.403921568627451,0.741176470588235}
\definecolor{color5}{rgb}{0.549019607843137,0.337254901960784,0.294117647058824}
\definecolor{color3}{rgb}{0.83921568627451,0.152941176470588,0.156862745098039}
\definecolor{color0}{rgb}{0.12156862745098,0.466666666666667,0.705882352941177}

\begin{axis}[
legend cell align={left},
legend style={font=\scriptsize,at={(0.5,0.5)}, anchor=south, draw=white!80.0!black},
tick align=inside,
tick pos=left,
x grid style={white!69.01960784313725!black},
xmajorgrids,
xmin=0, xmax=840,
xtick style={color=black},
y grid style={white!69.01960784313725!black},
ymajorgrids,
ymin=0, ymax=810,
ytick style={color=black},
hide y axis,
hide x axis,
legend columns=4
]

 \addplot [solid, color0]
 table [row sep=crcr] {%
 -1 -1\\
 -2 -2\\
 };
 \addlegendentry{\gls{gbf}}

 \addplot [solid, color1]
 table [row sep=crcr] {%
 -1 -1\\
 -2 -2\\
 };
 \addlegendentry{\gls{cbf}}

 \addplot [solid, color2]
 table [row sep=crcr] {%
 -1 -1\\
 -2 -2\\
 };
 \addlegendentry{\gls{fmbf}}

 \addplot [solid, color3]
 table [row sep=crcr] {%
 -1 -1\\
 -2 -2\\
 };
 \addlegendentry{\gls{smbf}}

\addplot [solid, black, thick]
table [row sep=crcr] {%
-1 -1\\
-2 -2\\
};
\addlegendentry{1 layer \gls{tmrs}}

\addplot [dashed, black, thick]
table [row sep=crcr] {%
-1 -1\\
-2 -2\\
};
\addlegendentry{4 layer \gls{pmrs}}
\end{axis}

\end{tikzpicture}
    \\
    \subfigure[\gls{cdf} of the \gls{ul} \gls{sinr}]{
      \setlength\fwidth{0.30\columnwidth}
      \setlength\fheight{0.25\columnwidth}
      \input{./figures/bf-comparison/cdf_sinr_ul_rlcAm=False_interPacketInterval=1500_harq=False.tex}
    \label{fig:bfulsinr}
    }
    \subfigure[\gls{cdf} of the \gls{dl} \gls{sinr}]{
      \setlength\fwidth{0.30\columnwidth}
      \setlength\fheight{0.25\columnwidth}
      \input{./figures/bf-comparison/cdf_sinr_dl_rlcAm=False_interPacketInterval=1500_harq=False.tex}
    \label{fig:bfdlsinr}
    }
    \subfigure[\gls{cdf} of the \gls{ul} \gls{bler}]{
      \setlength\fwidth{0.30\columnwidth}
      \setlength\fheight{0.25\columnwidth}
      \input{./figures/bf-comparison/cdf_bler_ul_rlcAm=False_interPacketInterval=1500_harq=False.tex}
    \label{fig:bfulbler}
    }
    \caption{Comparison of the different \gls{bf} schemes.}
    \vspace{-0.4cm}
\end{figure}

We compare the \gls{bf} schemes discussed in Section \ref{sec:bfdesign}. To clearly highlight their impact on the physical layer, we use \gls{rlc}-\gls{um} (i.e., without \gls{rlc} retransmissions), disable the \gls{harq} retransmissions at the \gls{mac} layer, and use a low-traffic application in the UEs. This minimizes the difference between the statistics of the \gls{sinr} and \gls{bler} perceived by the upper layers and the random distribution that generates these values at the channel model.

The low rate application is a constant traffic generator for downlink and uplink that produces a packet of 1500 bytes every 1500 $\mu$s, in each \gls{ue}. Roughly speaking, when the \gls{mcs} coding rate is greater than 3.64 bits per subcarrier, the 3300 subcarriers can carry a full packet in a single \gls{ofdm} symbol. This means that, in every slot of $250$ $\mu$s, the scheduler receives either a demand for at least $\sim$ 14 symbols (i.e., each of 7 \glspl{ue} requests one downlink and one uplink symbol), or none, in a regular pattern repeating every 6 slots. The scheduler has thus plenty of \glspl{rb} to satisfy the traffic request, and, as discussed, there are no retransmissions. This makes it possible to probe the channel and \gls{bf} scheme at a constant rate, and to measure the combined effect of the channel condition and \gls{bf} scheme on physical layer performance metrics.

We represent the received \gls{ul} \gls{sinr} \gls{cdf} for all transmission allocations in the simulation in Fig. \ref{fig:bfulsinr}. We compare 1 layer (solid) and 4 layer (dashed) cases. For the 1 layer case, we use the \gls{tmrs} without \gls{sdma} capabilities that was implemented in the previous versions of the ns-3 \gls{mmwave} module, with either \gls{gbf} or \gls{cbf} (\gls{mmse} \gls{bf} is a multi-layer technique and would have no effect in the single layer case, behaving exactly as \gls{cbf}). For the 4 layer case we consider the \gls{pmrs}, so that all allocations use the specified \gls{bf} scheme (i.e., the \gls{mmse} schemes never fall back to \gls{cbf}, as discussed in Sec.~\ref{sec:bfdesign}). We compare 4-layer \gls{gbf}, \gls{cbf}, \gls{fmbf}, \gls{smbf}. Since there is no self-interference, in the 1-layer case the \gls{sinr} is the same as the \gls{snr}, which is better with \gls{cbf} than with \gls{gbf}, consistently with our discussion in Sec.~\ref{sec:bfdesign}. Moreover, if we introduce 4 layers, but adopt any of the two single-layer \gls{bf} schemes (\gls{gbf} or \gls{cbf}), the \gls{sinr} drops significantly, with frequent -20~dB events. This confirms that the single-layer \gls{bf} schemes do not perform well and \textit{the use of multi-layer specific \gls{bf} is necessary}. Adopting the \gls{fmbf} scheme improves the \gls{sinr} by a significant margin, but does not fully compensate the interference. As designed, the \gls{smbf} scheme does remove almost all interference, and its \gls{sinr} \gls{cdf} is nearly identical to that of the single-layer \gls{cbf} case. 

We represent the received \gls{dl} \gls{sinr} \gls{cdf} in Fig. \ref{fig:bfdlsinr}. The main difference with the \gls{ul} case is that in \gls{dl} the desired and interfering signals at each UE arrive through the same pathloss, and -20 dB \gls{sinr} outages with 4-layer \gls{cbf} rarely happen. On the other hand, the gap between the higher range of \glspl{sinr} achieved with 4-layer \gls{cbf} and 4-layer \gls{mmse} \gls{bf} schemes is wider than in \gls{ul}. Again the \gls{smbf} scheme removes virtually all interference and achieves an \gls{sinr} distribution akin to that of a 1-layer \gls{cbf} scenario; whereas the \gls{fmbf} scheme achieves an intermediate \gls{sinr} improvement.

Finally we depict the instantaneous \gls{bler} \gls{cdf} for all \gls{ul} transmissions in Fig. \ref{fig:bfulbler}. Generally speaking, the \gls{bler} distribution is almost a step function: In each transmission, the \gls{cqi} feedback is used to select the \gls{mcs} such that the \gls{bler} would be $10^{-2}$ \textit{if the reported channel stayed the same}. Therefore, we can define a ``\gls{cqi} outage" as the event that, at the moment of transmission, the channel has become much worse compared to when the \gls{cqi} was reported, and the instantaneous \gls{bler} is $\simeq 1$. Fig.~\ref{fig:bfulbler} shows that the instantaneous \gls{bler} is dominated by such outages, where most transmissions experience either \gls{bler}$\leq 10^{-2}$ or \gls{bler}$=1$. The complement of this outage probability corresponds to the height of the flat region of the \gls{cdf} curves. As we can see, 4-layer \gls{gbf} and \gls{cbf} have a much larger outage probability (lower step in the \gls{bler} \gls{cdf}) and result in more severe \gls{bler} in the system. Again, we see that \gls{smbf} behaves almost as a 1-layer \gls{cbf} situation, with the \gls{fmbf} scheme in-between these two cases. We do not depict the \gls{dl} \gls{bler} \gls{cdf} due to space constraints, as its insights were identical.

In summary, the \gls{bf} comparison shows that \gls{smbf} is necessary in the \gls{mumimo} 4-layer implementation in order to ensure that the physical layer achieves the same \glspl{sinr} and \gls{bler} as in the 1-layer \gls{mmwave} system with \gls{cbf}. Moreover, the main differences between all \gls{bf} schemes in this paper are that \gls{gbf} performs worse than \gls{cbf} for any number of layers and that \gls{fmbf} offers some performance improvements without necessitating as much channel estimation overhead as \gls{smbf}.

\subsection{Cross-layer \gls{bf} and Scheduling Interactions}
\label{sec:bf-sched-results}

Next, we compare the performance of the scheduling algorithms introduced in Section \ref{sec:schdesign}. To highlight the scheduler behavior with respect to the offered traffic, once again we use \gls{rlc}-\gls{um} and disable the \gls{harq} retransmissions at the \gls{mac} layer. However, differently from the previous section, we use high-traffic applications in the UEs to emphasize the effect that the scheduler has on the system performance.

Since we have already determined the best \gls{bf} scheme for each number of layers, we consider four scenarios: the \gls{tmrs} 1-layer scheduler with \gls{cbf}, our proposed \gls{pmrs} for both the 1-layer \gls{cbf} and 4-layer \gls{smbf} configurations, and our proposed \gls{amrs} for 4-layers with \gls{smbf}. \gls{pmrs} is designed for use with multiple layers, but since it forces all allocations to be of the same size in \gls{tdma}, its behavior when applied to the 1-layer case differs slightly from that of \gls{tmrs}. For this reason we included an observation of the behavior of \gls{pmrs} in the 1-layer case as well. \gls{amrs}, on the other hand, behaves exactly like \gls{tmrs} if invoked on a 1-layer problem. Recall also that \gls{amrs} may assign allocations so that two overlapping transmissions do not start at the same time, and that in these events the \gls{smbf} scheme falls back to the behavior of a \gls{cbf} scheme (see Fig.~\ref{fig:bfschedconflict}).

The high-rate application is a constant bit rate source that generates a packet of 1500 bytes every 150 $\mu$s in each \gls{ue}, with a symmetric traffic in uplink and downlink. Recall that such packet can be sent with a single \gls{ofdm} symbol (with 3300 subcarriers) if the \gls{mcs} coding rate is greater than 3.64 bits/symbol. Overall, the traffic requests received by the scheduler saturate the \glspl{rb} of the 1-layer case. Indeed, for every slot of $250$ $\mu$s, the scheduler always receives requests for at least $\sim$~23 symbols. 
In the 1-layer case there are $12$ data symbols per slot, which are not enough to allocate all the demand. In the 4-layer case, there are $12\times4$ available data symbols, i.e.,  enough resources if most channels can support \glspl{mcs} with rates greater than 1.82 bits/symbol.

\begin{figure}[t]
    \centering
    \subfigure[Average \gls{bler}]{
      \setlength\fwidth{0.45\columnwidth}
      \setlength\fheight{0.3\columnwidth}
\begin{tikzpicture}

  \definecolor{color1}{rgb}{1,0.498039215686275,0.0549019607843137}
  \definecolor{color2}{rgb}{0.172549019607843,0.627450980392157,0.172549019607843}
  \definecolor{color4}{rgb}{0.580392156862745,0.403921568627451,0.741176470588235}
  \definecolor{color5}{rgb}{0.549019607843137,0.337254901960784,0.294117647058824}
  \definecolor{color3}{rgb}{0.83921568627451,0.152941176470588,0.156862745098039}
  \definecolor{color0}{rgb}{0.12156862745098,0.466666666666667,0.705882352941177}

\begin{axis}[
  ybar,
tick align=inside,
tick pos=both,
x grid style={white!69.01960784313725!black},
xtick style={color=black},
xtick={0,1,2,3},
xticklabels={\gls{tmrs}, \gls{pmrs}-1L, \gls{pmrs}-4L, \gls{amrs}},
y grid style={white!69.01960784313725!black},
ylabel={BLER},
ymin=0, ymax=0.5,
ytick style={color=black},
ytick={0,0.1,0.2,0.3,0.4,0.5},
yticklabels={0.0,0.1,0.2,0.3,0.4,0.5},
bar width=15pt,
xmin=0, xmax=3,
enlarge x limits=0.2,
legend columns=2, 
legend style={font=\tiny,at={(0.02,0.98)}, anchor=north west, draw=white!80.0!black},
label style={font=\tiny},
tick label style={font=\tiny} 
]

\addlegendimage{ybar,ybar legend, color0, fill=color0,fill opacity=0.4};
\addlegendentry{DL}

\addlegendimage{ybar,ybar legend, color1, fill=color1,fill opacity=0.4, postaction={pattern=north east lines}};
\addlegendentry{UL}

\addplot[color0, fill=color0, fill opacity=0.4] coordinates {(0,0.123515226065967) (1,0.124066877923332) (2, 0.159321238785321) (3,0.161688326366552)};

\addplot[color1, fill=color1, fill opacity=0.4, postaction={pattern=north east lines}] coordinates {(0,0.242682975360048) (1,0.136747518024729) (2,0.136730253662689) (3,0.390675178226012)};

%

%
\end{axis}
\end{tikzpicture}
    \label{fig:schedbler}
    }
    \hspace{.01in}
    \subfigure[Throughput]{
      \setlength\fwidth{0.45\columnwidth}
      \setlength\fheight{0.3\columnwidth}
\begin{tikzpicture}

  \definecolor{color1}{rgb}{1,0.498039215686275,0.0549019607843137}
  \definecolor{color2}{rgb}{0.172549019607843,0.627450980392157,0.172549019607843}
  \definecolor{color4}{rgb}{0.580392156862745,0.403921568627451,0.741176470588235}
  \definecolor{color5}{rgb}{0.549019607843137,0.337254901960784,0.294117647058824}
  \definecolor{color3}{rgb}{0.83921568627451,0.152941176470588,0.156862745098039}
  \definecolor{color0}{rgb}{0.12156862745098,0.466666666666667,0.705882352941177}

  \begin{axis}[
    ybar,
  tick align=inside,
  tick pos=both,
  x grid style={white!69.01960784313725!black},
  xtick style={color=black},
  xtick={0,1,2,3},
xticklabels={TMRS, PMRS-1L, PMRS-4L, AMRS},
  y grid style={white!69.01960784313725!black},
  ylabel={Throughput [Mbps]},
  ymin=0, ymax=60e1,
  ytick style={color=black},
  bar width=15pt,
  xmin=0, xmax=3,
  enlarge x limits=0.2,
  legend columns=2,
  legend style={font=\tiny,at={(0.02,0.98)}, anchor=north west, draw=white!80.0!black},
    label style={font=\tiny},
    tick label style={font=\tiny} 
  ]

  \addlegendimage{ybar,ybar legend, color0, fill=color0,fill opacity=0.4};
  \addlegendentry{DL}

  \addlegendimage{ybar,ybar legend, color1, fill=color1,fill opacity=0.4, postaction={pattern=north east lines}};
  \addlegendentry{UL}

  \addplot[color0, fill=color0, fill opacity=0.4] coordinates {(0,334.253935) (1,297.716900) 
  (2,424.865305) (3,410.301300)};
  
  \addplot[color1, fill=color1, fill opacity=0.4, postaction={pattern=north east lines}] coordinates {(0,182.657610) (1,228.338775) 
  (2,459.142755) (3,252.1695715)};

  %
  
\end{axis}

\end{tikzpicture}
    \label{fig:schedulDataRx}
    }
    \caption{Comparison of the different scheduling strategies.}
    \vspace{-0.4cm}
\end{figure}
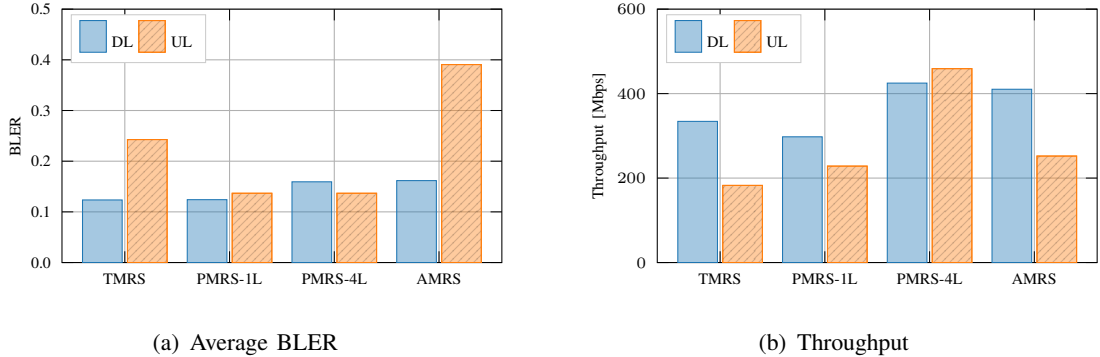

Figure \ref{fig:schedbler} reports the average \gls{dl} and \gls{ul} \gls{bler} for the four scheduler-\gls{bf} pairs discussed above. As can be seen, \gls{amrs} displays a high \gls{ul} \gls{bler} because it does not fully take advantage of the \gls{smbf} technique. The problem is more severe in \gls{ul} because, as discussed in the previous section, the pathloss leads to more severe \gls{sinr} drops (outages) in this direction than in \gls{dl}. The \gls{bler} of \gls{pmrs} with 4 layers is comparable to that of \gls{tmrs} and \gls{pmrs} with 1 layer, which is also consistent with the \gls{sinr} plots discussed in the previous section.

Figure \ref{fig:schedulDataRx} depicts the throughput, defined as total data received divided by total simulation duration. Since the nominal application rate is $\frac{7 \times 1500 \times 8}{150 \times 10^{-6}}$ bit/s, the maximum throughput is  $560$~Mbps. For \gls{tmrs}, we see that the delivered throughput is around 330~Mbps in \gls{dl} and 180~Mbps in \gls{ul}, with significant asymmetry and much lower value than the offered traffic. This is consistent with the fact that the offered traffic greatly exceeds the number of data symbols of the 1-layer frame even with the best \gls{mcs}. The same observation holds for the padding scheduler with 1-layer and \gls{cbf}, although its \gls{dl}/\gls{ul} traffic is better balanced. 
In the 4-layer cases, the resources are not saturated by the source traffic, and the throughput with \gls{pmrs} exceeds 420~Mbps in \gls{dl} and 450~Mbps in \gls{ul}. This shows the main advantage of incorporating \gls{sdma} \gls{mumimo} into \gls{mmwave} networks, i.e., \textit{an increase in the number of available \glspl{rb} by a factor of $N_{layers}$ allows the network to support much more traffic.} Particularly our simulation shows a delivered traffic that is $2\times$ the capacity of the single-layer frame. As for \gls{amrs}, we see that it also supported over 410~Mbps of \gls{dl} traffic successfully, but it only delivered around 250~Mbps of \gls{ul} traffic. This is consistent with the observation that the \gls{ul} \gls{bler} is high and the \gls{sinr} can suffer significant outages when \gls{amrs} does not ensure that all layers start their transmissions at the same time, causing the \gls{smbf} scheme to fall back to \gls{cbf} behavior. Therefore, despite the higher efficiency in the resource allocation, \gls{amrs} may perform worse than \gls{pmrs}.

This section shows that the implementation of multi-layer SDMA \gls{mumimo} in the \gls{mmwave} cellular system can greatly increase its capacity. However, to do so, attention must be paid to the inter-beam interference and scheduling conflict. In the previous section we had established that \gls{cbf} in a multi-layer setting suffers occasional severe outages in \gls{ul}. In this section we also show that this issue extends to \gls{smbf} used in combination with \gls{amrs}, since in this case the \gls{bf} scheme must occasionally fall back to \gls{cbf} behavior. 

\begin{figure}
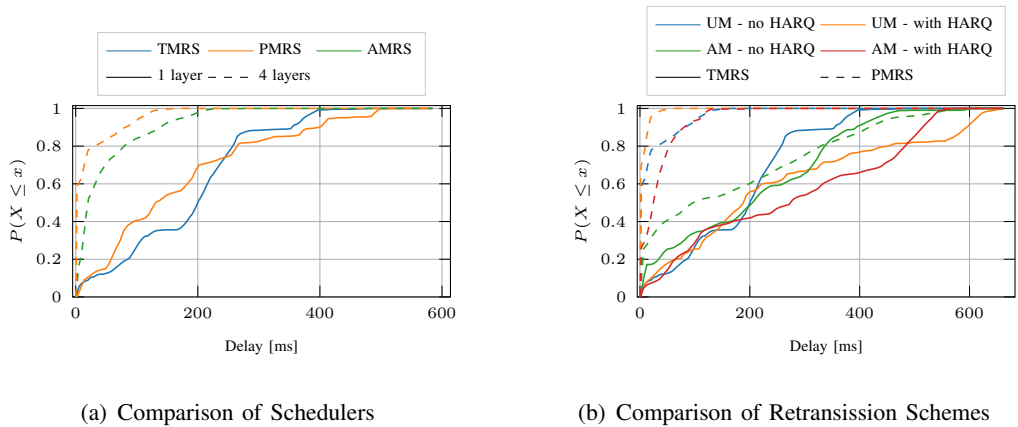

    \centering
    \subfigure[Comparison of Schedulers]{
      \setlength\fwidth{0.4\columnwidth}
      \setlength\fheight{0.25\columnwidth}
      \input{./figures/sched-comparison/cdf_delay_ul_rlcAm=False_interPacketInterval=150_harq=False.tex}
    \label{fig:schedulDelay}
    }
    \hspace{.1in}
    \subfigure[Comparison of Retransission Schemes]{
      \setlength\fwidth{0.4\columnwidth}
      \setlength\fheight{0.25\columnwidth}
      \input{./figures/sched-comparison/cdf_delay_ul_interPacketInterval=150_retx.tex}
    \label{fig:schedretxComp}
    }
    \caption{Delay in \gls{ul}}
    \vspace{-0.4cm}
\end{figure}

\subsection{Delay and Retransmissions}

Capacity is not the only relevant network performance indicator. Some applications such as video streaming or eHealth are time constrained, whereas other applications such as smart grid and machine-type communications require strong reliability. In this section we leverage the end-to-end nature of our simulator to study the relation between the available retransmission schemes and delay in \gls{sdma} \gls{mumimo} \gls{mmwave} \gls{nr} networks. We note that delay measurements (which are taken at the \gls{pdcp} layer in the 3GPP stack) in simulations are affected by the so-called ``survivor bias," i.e., packets that are not delivered do not get their delay measured. For this reason unreliable transmission modes tend to display shorter delay statistics on the fewer packets successfully arriving, whereas the queuing and retransmission timeouts of reliable modes add to their total delay on a greater number of received packets. The \gls{rlc} retransmission mode (i.e., \gls{rlc} \gls{am}) provides reliability on a much larger time scale than the \gls{mac} \gls{harq} mechanism. The \gls{rlc} ``reordering timeout" is 10~ms, whereas the \gls{harq} scheme retransmits immediately after a \gls{nack} is received. Since \gls{ul} control information is processed at the end of every slot in our simulation, this makes the \gls{harq} retransmission time less than 1 slot period or $250~\mu$s. For this reason we expect that \gls{rlc}-\gls{am} will dominate the increase in delay caused by retransmissions.

Figure \ref{fig:schedulDelay} compares the delay \glspl{cdf} for different schedulers under the no retransmission configuration (\gls{rlc} \gls{um} without \gls{harq}), with the high traffic \gls{udp} application presented in the previous section. We use the \gls{cdf} instead of bar plots to highlight the inverted-L shape of delay \glspl{cdf} when most traffic is successfully delivered in the 4-layer configuration. With this, more than 75\% of packets are received in under 20~ms, and more than 90\% of packets are received under 100~ms. In a deadline constrained application, such as for example video with one frame every 20~ms, this means that 75\% of the frames would be received on time and be displayed on screen (with no buffering). 
Turning our attention to the differences between \gls{amrs} and \gls{pmrs}, the latter indeed guarantees a 10~ms deadline with probability 80\% and a 100~ms deadline with probability 95\%, which is much better than the deadline guarantees offered by \gls{amrs}. The 1-layer schemes, both \gls{tmrs} and 1-layer-\gls{pmrs}, do not display an inverse-L shaped \gls{cdf} because the network capacity is exceeded by the applications. Instead, the delay \gls{cdf} with 1-layer is roughly linear as many packets accumulated long times in the queue waiting to be transmitted.

Figure \ref{fig:schedretxComp} displays the delay \gls{cdf} using all four possible retransmission configurations for \gls{tmrs} and 4-layer \gls{pmrs}. Again we adopt the \gls{udp} high traffic application. Since the offered throughput exceeds the resources of the 1-layer frame the delays with \gls{tmrs} present an almost-linear slope which is dominated by queue waiting time. On the other hand the \gls{pmrs} cases present two-slope inverse-L shapes that are mostly driven by outages and retransmissions. The lowest delay 80\%-tile is achieved by the \gls{rlc}-\gls{um} with \gls{harq} \gls{pmrs} configuration, followed closely by the \gls{rlc}-\gls{um} without \gls{harq} \gls{pmrs} configuration. It appears that \gls{harq} retransmissions help improve delay, which suggests that their contribution to improve reliability compensates the small delays incurred by \gls{harq} retransmissions. 
The \glspl{cdf}s for \gls{pmrs} with \gls{rlc} \gls{am} exhibit a very different behavior with or without \gls{harq} retransmissions. Without \gls{harq}, multiple \gls{am} restransmissions are needed, where each retransmission adds over 10~ms to the packet delivery delay. On the contrary, with \gls{harq} most retransmissions take place at the \gls{mac} layer, with a short round-trip time, and \gls{rlc} only needs to compensate for occasional \gls{harq} failures. It is noteworthy that the delay \gls{cdf} for the \gls{rlc} \gls{am} without \gls{harq} padding configuration looks similar to the 1-layer curves, which suggests that \gls{rlc} retransmission queues are growing without bounds in this scenario. Regarding the differences in behavior between different retransmission configurations for \gls{tmrs}, it seems that resource occupation dominates the delay since the 1-layer frame capacity is exceeded. That is to say, the \gls{rlc} \gls{um} without \gls{harq} 1-layer configuration does not add any resource demands besides that of applications, thus alleviating the queues, whereas the \gls{rlc} \gls{am} with \gls{harq} configuration adds resource demands to the scheduler on top of the demands already presented by the fresh packets, making the queues grow even longer and the delay worsen.

The main conclusion of this section is that, since delay is strongly related to resource availability and queuing, the use of \gls{sdma} greatly increases the number of available resource blocks, permitting the schedulers to support larger traffic demands with low delay. Among the schedulers, \gls{pmrs} offers an improved delay profile with respect to \gls{amrs}, but both are able to offer under 20~ms delay to a high percentage of the traffic. For intuitive reference, a video at 50 frames per second displays one frame every 20~ms, so this result is of the same order of magnitude as real-time multimedia applications. To introduce reliability, \gls{harq} should be activated always first before considering the use of the \gls{rlc} \gls{am} mode, as following the opposite order would cause too many retransmissions and delay at the \gls{rlc} level. 

\subsection{Throughput vs Delay}

This subsection further extends the scheduler comparison by considering joint throughput and delay results using the ``full retransmission'' scheme, i.e., the \gls{rlc} \gls{am} mode with \gls{harq}, versus the scenario ``without retransmissions'' consisting in using the \gls{rlc} \gls{um} without \gls{harq} retransmissions. As in the previous section, we consider a high-rate \gls{udp} application with 150 $\mu$s inter-packet-interval and focus on the delay and throughput. We compare the default \gls{tmrs} in the ns-3 \gls{mmwave} module with 1 layer versus our \gls{pmrs} and \gls{amrs} with 4 layers.

\begin{figure}
    \centering
    \subfigure[Uplink]{
      \setlength\fwidth{0.45\columnwidth}
      \setlength\fheight{0.3\columnwidth}
\begin{tikzpicture}
\pgfplotsset{every tick label/.append style={font=\scriptsize}}

\definecolor{color1}{rgb}{1,0.498039215686275,0.0549019607843137}
\definecolor{color2}{rgb}{0.172549019607843,0.627450980392157,0.172549019607843}
\definecolor{color4}{rgb}{0.580392156862745,0.403921568627451,0.741176470588235}
\definecolor{color5}{rgb}{0.549019607843137,0.337254901960784,0.294117647058824}
\definecolor{color3}{rgb}{0.83921568627451,0.152941176470588,0.156862745098039}
\definecolor{color0}{rgb}{0.12156862745098,0.466666666666667,0.705882352941177}

\begin{axis}[
legend columns=3,
xmode=log,
xlabel style={font=\scriptsize},
xlabel={Delay [ms]},
xmajorgrids,
xmin=7, xmax=350,
xtick style={color=white!15!black},
ylabel style={font=\scriptsize},
ylabel={Throughput [Mbps]},
ymajorgrids,
ymin=100, ymax=500,
ytick style={color=white!15!black},
label style={font=\scriptsize},
tick label style={font=\scriptsize} 
]

\addplot[
  scatter,
  only marks,
  scatter src=explicit,
  scatter/classes={1={color0}, 2={color1}},
  mark=otimes,
  mark size=4,
  forget plot,
]
table[x=x,y=y, meta=class]{%
x                      y              class
275.946375845866 148.081415 1
188.828798402743 182.65761 2

};

\addplot[
  scatter,
  only marks,
  scatter src=explicit,
  scatter/classes={1={color0}, 2={color1}},
  forget plot,
  mark=10-pointed star,
  mark size=4,
]
table[x=x,y=y, meta=class]{%
x                      y              class
28.0083495614619 270.429785 1
8.96418240590719 459.142755 2
};

\addplot[
  scatter,
  only marks,
  scatter src=explicit,
  scatter/classes={1={color0}, 2={color1}},
  mark size=4,
]
table[x=x,y=y, meta=class]{%
x                      y              class
150.798441964601 165.31997 1
55.9157594215237 252.1695715 2
};


%
%
%


\end{axis}

\begin{axis}[
legend cell align={left},
legend style={font=\scriptsize,at={(0.5,1)}, anchor=south, draw=white!80.0!black},
tick align=inside,
tick pos=left,
x grid style={white!69.01960784313725!black},
xmajorgrids,
xmin=0, xmax=840,
xtick style={color=black},
y grid style={white!69.01960784313725!black},
ymajorgrids,
ymin=0, ymax=810,
ytick style={color=black},
hide y axis,
hide x axis,
legend columns=3,
]

\addplot[
  scatter,
  only marks,
  scatter src=explicit,
  mark size=4,
  mark=otimes,
  color=black,
]
table[x=x,y=y]{%
x                      y
-20 -22
};
\addlegendentry{\gls{tmrs}}

\addplot[
  scatter,
  only marks,
  scatter src=explicit,
  mark size=4,
  mark=10-pointed star,
  color=black,
]
table[x=x,y=y]{%
x                      y
-20 -21 
};
\addlegendentry{\gls{pmrs}}

\addplot[
  scatter,
  only marks,
  scatter src=explicit,
  mark size=4,
  color=black,
]
table[x=x,y=y]{%
x                      y
-20 -20 
};
\addlegendentry{\gls{amrs}}

\addplot[
  scatter,
  only marks,
  scatter src=explicit,
  mark size=4,
  color=color0,
]
table[x=x,y=y]{%
x                      y
-20 -20 
};
\addlegendentry{With retx}

\addplot[
  scatter,
  only marks,
  scatter src=explicit,
  mark size=4,
  color=color1,
]
table[x=x,y=y]{%
x                      y
-20 -20 
};
\addlegendentry{Without retx}
\end{axis}

\end{tikzpicture}
    \label{fig:schedscatterUl}
    }
    \hspace{.01in}
    \subfigure[Downlink]{
      \setlength\fwidth{0.45\columnwidth}
      \setlength\fheight{0.3\columnwidth}
\begin{tikzpicture}
\pgfplotsset{every tick label/.append style={font=\scriptsize}}

\definecolor{color1}{rgb}{1,0.498039215686275,0.0549019607843137}
\definecolor{color2}{rgb}{0.172549019607843,0.627450980392157,0.172549019607843}
\definecolor{color4}{rgb}{0.580392156862745,0.403921568627451,0.741176470588235}
\definecolor{color5}{rgb}{0.549019607843137,0.337254901960784,0.294117647058824}
\definecolor{color3}{rgb}{0.83921568627451,0.152941176470588,0.156862745098039}
\definecolor{color0}{rgb}{0.12156862745098,0.466666666666667,0.705882352941177}

\begin{axis}[
legend columns=3,
xmode=log,
xlabel style={font=\scriptsize},
xlabel={Delay [ms]},
xmajorgrids,
xmin=7, xmax=150,
xtick style={color=white!15!black},
ylabel style={font=\scriptsize},
ylabel={Throughput [Mbps]},
ymajorgrids,
ymin=280, ymax=500,
ytick style={color=white!15!black},
label style={font=\scriptsize},
tick label style={font=\scriptsize} 
]

\addplot[
  scatter,
  only marks,
  scatter src=explicit,
  scatter/classes={1={color0}, 2={color1}},
  mark=otimes,
  mark size=4,
  forget plot,
]
table[x=x,y=y, meta=class]{%
x                      y              class
118.684065037563 344.043235 1
96.2057991834283 334.253935 2

};

\addplot[
  scatter,
  only marks,
  scatter src=explicit,
  scatter/classes={1={color0}, 2={color1}},
  forget plot,
  mark=10-pointed star,
  mark size=4,
]
table[x=x,y=y, meta=class]{%
x                      y              class
8.73764207931606 302.035665 1
8.94045855047831 424.865305 2
};

\addplot[
  scatter,
  only marks,
  scatter src=explicit,
  scatter/classes={1={color0}, 2={color1}},
  mark size=4,
]
table[x=x,y=y, meta=class]{%
x                      y              class
61.0805252499937 464.44641 1
23.6781017340634 410.3013 2
};


%
%
%


\end{axis}

\begin{axis}[
legend cell align={left},
legend style={font=\scriptsize,at={(0.5,1)}, anchor=south, draw=white!80.0!black},
tick align=inside,
tick pos=left,
x grid style={white!69.01960784313725!black},
xmajorgrids,
xmin=0, xmax=840,
xtick style={color=black},
y grid style={white!69.01960784313725!black},
ymajorgrids,
ymin=0, ymax=810,
ytick style={color=black},
hide y axis,
hide x axis,
legend columns=3,
]

\addplot[
  scatter,
  only marks,
  scatter src=explicit,
  mark size=4,
  mark=otimes,
  color=black,
]
table[x=x,y=y]{%
x                      y
-20 -22
};
\addlegendentry{\gls{tmrs}}

\addplot[
  scatter,
  only marks,
  scatter src=explicit,
  mark size=4,
  mark=10-pointed star,
  color=black,
]
table[x=x,y=y]{%
x                      y
-20 -21 
};
\addlegendentry{\gls{pmrs}}

\addplot[
  scatter,
  only marks,
  scatter src=explicit,
  mark size=4,
  color=black,
]
table[x=x,y=y]{%
x                      y
-20 -20 
};
\addlegendentry{\gls{amrs}}

\addplot[
  scatter,
  only marks,
  scatter src=explicit,
  mark size=4,
  color=color0,
]
table[x=x,y=y]{%
x                      y
-20 -20 
};
\addlegendentry{With retx}

\addplot[
  scatter,
  only marks,
  scatter src=explicit,
  mark size=4,
  color=color1,
]
table[x=x,y=y]{%
x                      y
-20 -20 
};
\addlegendentry{Without retx}
\end{axis}

\end{tikzpicture}
    \label{fig:schedscatterDl}
    }
    \caption{Comparison of the different scheduling strategies.}
    \vspace{-0.4cm}
\end{figure}
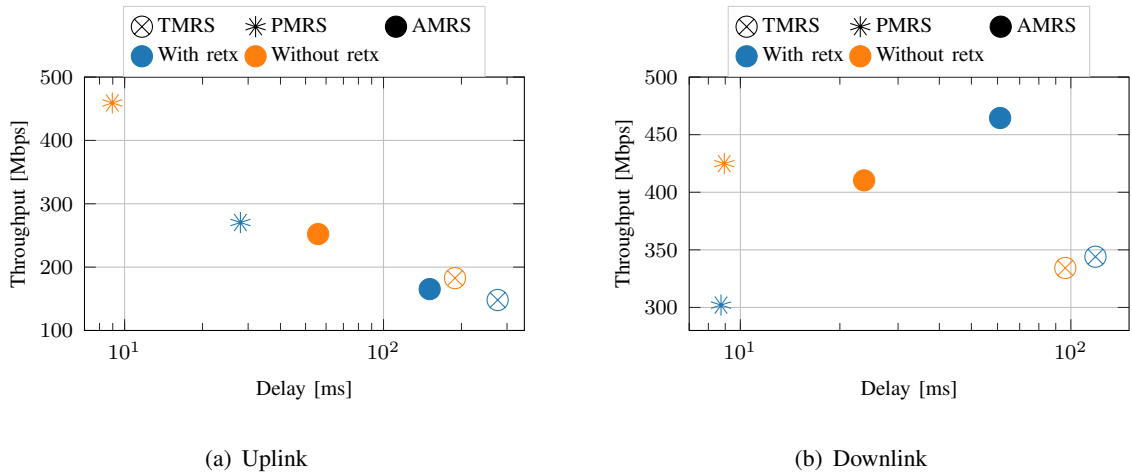

Figure \ref{fig:schedscatterUl} reports the mean \gls{ul} delay vs throughput. Each point in the scatter cloud corresponds to one possible system configuration, with the best configuration corresponding to the top left corner, i.e., the highest throughput with the lowest delay. Recalling that the \textit{offered} traffic is $7\times 1500\times 8 /150 \times 10^{-6}=560$~Mbps, we note that the 4-layer padding scheduler without retransmissions is the only one to deliver almost all the traffic. Surprisingly, activating the \gls{rlc} \gls{am} mode with \gls{harq} reduces the throughput, which means that the additional \gls{rb} demand of the retransmissions overweighs the benefit of increased reliability. Since the offered traffic greatly exceeds the capacity of a 1-layer case, \gls{tmrs} with 1-layer  displays large delays (waiting in queues) and low throughput. This figure focuses to the \gls{ul} performance, in which \gls{amrs} suffers occasional outage problems, and hence its throughput and delay are much worse than \gls{pmrs}.

Figure \ref{fig:schedscatterDl}, instead, reports the same metrics for the \gls{dl} traffic. The major difference with the \gls{ul} case is that now \gls{amrs} performs much better. Indeed, with retransmissions enabled, \gls{amrs} displays the highest throughput at the cost of a slightly higher delay (due to the \gls{rlc} \gls{am} retransmission timer). On the other hand, \gls{pmrs} displays a significant drop in throughput with retransmissions versus the case without them. In this case, indeed, the padding brings the resource occupation close to saturation and the retransmissions, which increase the resource demand, saturate the capacity and cause a net decrease in throughput. \gls{amrs}, on the other hand, is not affected by this issue (at least not yet, with this source rate), as it is more efficient in allocating resources.

The results in this section highlight the importance of a full-stack, end-to-end performance evaluation. Indeed, the evaluation of the BLER and SINR in Sec.~\ref{sec:bf-sched-results} seemed to suggest that \gls{amrs} always performed equal or worse than \gls{pmrs}. However, in \gls{dl}, the \gls{bler} penalty of \gls{amrs} can be compensated using retransmissions, and, overall, the more efficient resource allocation yields an improved throughput. Conversely, since \gls{pmrs} wastes some frame resources, the activation of retransmissions worsens the cell saturation and penalizes its throughput instead of helping it. Nonetheless, in \gls{ul} \gls{amrs} severely underperforms \gls{pmrs}. This suggests that different scheduling principles could be adopted for the two directions.

\subsection{Performance with different traffic sources}
\label{sec:sources}

Finally, we compare the system performance under three different applications and transport layer configurations, and investigate the relation between the application traffic and the scheduler. The first two applications are those considered in Sec.~\ref{sec:bfcomparison} and Sec.~\ref{sec:bf-sched-results}, i.e., a constant bitrate source that generates a packet of 1500 bytes every 1500 or 150 $\mu$s. In this case, the transport layer is \gls{udp} (thus they will be referred to as \emph{\gls{udp} slow} and \emph{\gls{udp} fast}, respectively). Finally, we also profile the performance with a full buffer application that relies on \gls{tcp} at the transport layer, to adjust the offered traffic to the maximum supported by the network. We consider retransmissions in the \gls{rlc} and \gls{mac} layers to obtain similar reliability in the applications over \gls{udp} as in the application over \gls{tcp}.

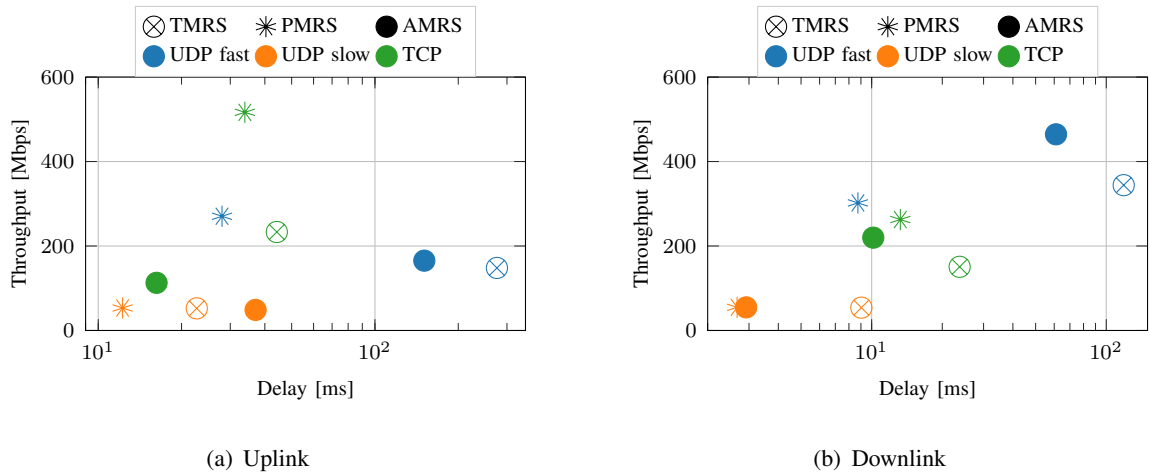
\begin{figure}
    \centering
    \subfigure[Uplink]{
      \setlength\fwidth{0.45\columnwidth}
      \setlength\fheight{0.3\columnwidth}
\begin{tikzpicture}
\pgfplotsset{every tick label/.append style={font=\scriptsize}}

\definecolor{color1}{rgb}{1,0.498039215686275,0.0549019607843137}
\definecolor{color2}{rgb}{0.172549019607843,0.627450980392157,0.172549019607843}
\definecolor{color4}{rgb}{0.580392156862745,0.403921568627451,0.741176470588235}
\definecolor{color5}{rgb}{0.549019607843137,0.337254901960784,0.294117647058824}
\definecolor{color3}{rgb}{0.83921568627451,0.152941176470588,0.156862745098039}
\definecolor{color0}{rgb}{0.12156862745098,0.466666666666667,0.705882352941177}

\begin{axis}[
legend columns=3,
xmode=log,
xlabel style={font=\scriptsize},
xlabel={Delay [ms]},
xmajorgrids,
xmin=9, xmax=350,
xtick style={color=white!15!black},
ylabel style={font=\scriptsize},
ylabel={Throughput [Mbps]},
ymajorgrids,
ymin=0, ymax=600,
ytick style={color=white!15!black},
label style={font=\scriptsize},
tick label style={font=\scriptsize} 
]

\addplot[
  scatter,
  only marks,
  scatter src=explicit,
  scatter/classes={1={color0}, 2={color1}, 3={color2}},
  mark=otimes,
  mark size=4,
  forget plot,
]
table[x=x,y=y, meta=class]{%
x                      y              class
275.946375845866 148.081415 1
22.6998788352737 52.34072 2
44.188010245138 233.156498 3
};

\addplot[
  scatter,
  only marks,
  scatter src=explicit,
  scatter/classes={1={color0}, 2={color1}, 3={color2}},
  forget plot,
  mark=10-pointed star,
  mark size=4,
]
table[x=x,y=y, meta=class]{%
x                      y              class
28.0083495614619 270.429785 1
12.2439280373733 53.251855 2
33.8720792069212 516.502764 3
};

\addplot[
  scatter,
  only marks,
  scatter src=explicit,
  scatter/classes={1={color0}, 2={color1}, 3={color2}},
  mark size=4,
]
table[x=x,y=y, meta=class]{%
x                      y              class
150.798441964601 165.31997 1
37.027599172488 48.677555 2
16.2454111023011 112.832777 3
};


%
%
%


\end{axis}

\begin{axis}[
legend cell align={left},
legend style={font=\scriptsize,at={(0.5,1)}, anchor=south, draw=white!80.0!black},
tick align=inside,
tick pos=left,
x grid style={white!69.01960784313725!black},
xmajorgrids,
xmin=0, xmax=840,
xtick style={color=black},
y grid style={white!69.01960784313725!black},
ymajorgrids,
ymin=0, ymax=810,
ytick style={color=black},
hide y axis,
hide x axis,
legend columns=3,
]

\addplot[
  scatter,
  only marks,
  scatter src=explicit,
  mark size=4,
  mark=otimes,
  color=black,
]
table[x=x,y=y]{%
x                      y
-20 -22
};
\addlegendentry{\gls{tmrs}}

\addplot[
  scatter,
  only marks,
  scatter src=explicit,
  mark size=4,
  mark=10-pointed star,
  color=black,
]
table[x=x,y=y]{%
x                      y
-20 -21 
};
\addlegendentry{\gls{pmrs}}

\addplot[
  scatter,
  only marks,
  scatter src=explicit,
  mark size=4,
  color=black,
]
table[x=x,y=y]{%
x                      y
-20 -20 
};
\addlegendentry{\gls{amrs}}

\addplot[
  scatter,
  only marks,
  scatter src=explicit,
  mark size=4,
  color=color0,
]
table[x=x,y=y]{%
x                      y
-20 -20 
};
\addlegendentry{UDP fast}

\addplot[
  scatter,
  only marks,
  scatter src=explicit,
  mark size=4,
  color=color1,
]
table[x=x,y=y]{%
x                      y
-20 -20 
};
\addlegendentry{UDP slow}

\addplot[
  scatter,
  only marks,
  scatter src=explicit,
  mark size=4,
  color=color2,
]
table[x=x,y=y]{%
x                      y
-20 -20 
};
\addlegendentry{TCP}
\end{axis}

\end{tikzpicture}
    \label{fig:schedscatterUlApp}
    }
    \hspace{.1in}
    \subfigure[Downlink]{
      \setlength\fwidth{0.45\columnwidth}
      \setlength\fheight{0.3\columnwidth}
\begin{tikzpicture}
\pgfplotsset{every tick label/.append style={font=\scriptsize}}

\definecolor{color1}{rgb}{1,0.498039215686275,0.0549019607843137}
\definecolor{color2}{rgb}{0.172549019607843,0.627450980392157,0.172549019607843}
\definecolor{color4}{rgb}{0.580392156862745,0.403921568627451,0.741176470588235}
\definecolor{color5}{rgb}{0.549019607843137,0.337254901960784,0.294117647058824}
\definecolor{color3}{rgb}{0.83921568627451,0.152941176470588,0.156862745098039}
\definecolor{color0}{rgb}{0.12156862745098,0.466666666666667,0.705882352941177}

\begin{axis}[
legend columns=3,
xlabel style={font=\scriptsize},
xlabel={Delay [ms]},
xmajorgrids,
xmin=2, xmax=150,
xtick style={color=white!15!black},
ylabel style={font=\scriptsize},
ylabel={Throughput [Mbps]},
ymajorgrids,
ymin=0, ymax=600,
ytick style={color=white!15!black},
xmode=log,
label style={font=\scriptsize},
tick label style={font=\scriptsize} 
]

\addplot[
  scatter,
  only marks,
  scatter src=explicit,
  scatter/classes={1={color0}, 2={color1}, 3={color2}},
  mark=otimes,
  mark size=4,
  forget plot,
]
table[x=x,y=y, meta=class]{%
x                      y              class
118.684065037563 344.043235 1
9.03851033310803 54.017715 2
23.7142270896674 150.755171 3

};

\addplot[
  scatter,
  only marks,
  scatter src=explicit,
  scatter/classes={1={color0}, 2={color1}, 3={color2}},
  forget plot,
  mark=10-pointed star,
  mark size=4,
]
table[x=x,y=y, meta=class]{%
x                      y              class
8.73764207931606 302.035665 1
2.66946781869068 54.78581 2
13.250359863506 262.651234 3
};

\addplot[
  scatter,
  only marks,
  scatter src=explicit,
  scatter/classes={1={color0}, 2={color1}, 3={color2}},
  mark size=4,
]
table[x=x,y=y, meta=class]{%
x                      y              class
61.0805252499937 464.44641 1
2.91835095741966 54.69343 2
10.159532817922 219.78672 3
};


%
%
%


\end{axis}

\begin{axis}[
legend cell align={left},
legend style={font=\scriptsize,at={(0.5,1)}, anchor=south, draw=white!80.0!black},
tick align=inside,
tick pos=left,
x grid style={white!69.01960784313725!black},
xmajorgrids,
xmin=0, xmax=840,
xtick style={color=black},
y grid style={white!69.01960784313725!black},
ymajorgrids,
ymin=0, ymax=810,
ytick style={color=black},
hide y axis,
hide x axis,
legend columns=3,
]

\addplot[
  scatter,
  only marks,
  scatter src=explicit,
  mark size=4,
  mark=otimes,
  color=black,
]
table[x=x,y=y]{%
x                      y
-20 -22
};
\addlegendentry{\gls{tmrs}}

\addplot[
  scatter,
  only marks,
  scatter src=explicit,
  mark size=4,
  mark=10-pointed star,
  color=black,
]
table[x=x,y=y]{%
x                      y
-20 -21 
};
\addlegendentry{\gls{pmrs}}

\addplot[
  scatter,
  only marks,
  scatter src=explicit,
  mark size=4,
  color=black,
]
table[x=x,y=y]{%
x                      y
-20 -20 
};
\addlegendentry{\gls{amrs}}

\addplot[
  scatter,
  only marks,
  scatter src=explicit,
  mark size=4,
  color=color0,
]
table[x=x,y=y]{%
x                      y
-20 -20 
};
\addlegendentry{UDP fast}

\addplot[
  scatter,
  only marks,
  scatter src=explicit,
  mark size=4,
  color=color1,
]
table[x=x,y=y]{%
x                      y
-20 -20 
};
\addlegendentry{UDP slow}

\addplot[
  scatter,
  only marks,
  scatter src=explicit,
  mark size=4,
  color=color2,
]
table[x=x,y=y]{%
x                      y
-20 -20 
};
\addlegendentry{TCP}
\end{axis}

\end{tikzpicture}
    \label{fig:schedscatterDlApp}
    }
    \caption{Comparison of different applications.}
    \vspace{-0.4cm}
\end{figure}

Figure \ref{fig:schedscatterUlApp} represents the \gls{ul} delay vs throughput for all three applications and all three scheduling solutions. Since in the \gls{udp} slow application (in yellow) the offered traffic is much lower than the potential cell capacity, almost all source rate is successfully delivered by all schedulers (about 56~Mbps). In addition, \gls{pmrs} displays the lowest delay, followed by \gls{tmrs}, with \gls{amrs} offering the worst \gls{ul} delay. As discussed throughout the prior sections this is because of the occasional events where \gls{amrs} suffers deep \gls{sinr} outages in \gls{ul}. In the \gls{udp} fast application (in blue) the traffic sources offer $10\times$ more throughput, which is almost fully delivered using \gls{pmrs}. \gls{tmrs} and \gls{amrs} do not deliver all the \gls{ul} traffic for different reasons. While in \gls{tmrs} this is due to the limited resources of the 1-layer frame, in \gls{amrs} the reason is the high \gls{bler} due to the occasional outages. Since the \gls{udp} fast application does not adjust its transmission queues the delay in these two schedulers increases significantly. Finally, for the \gls{tcp} application (in green), \gls{pmrs} offers the best performance achieving about 560 Mbps. \gls{tmrs} has limited resources in the 1-layer frame and hence the throughput is less than half, but the delay is tolerable under 50~ms. Finally \gls{amrs} achieves a very low rate, which can be explained by the \gls{tcp} rate control responding too strongly to the occasional SINR outages, which produce packet losses that trigger the \gls{tcp} congestion control, reducing the transmission window.

Similarly, Fig. \ref{fig:schedscatterDlApp} represents the \gls{dl} delay vs throughput for all three applications and all three scheduling solutions. As in the previous figures, the main difference is that \gls{amrs} performs much better in \gls{dl} than in \gls{ul}. For the \gls{udp} slow application we still see that all the traffic is delivered, but the source rate is small. In \gls{dl} the delay is much lower and similar between the two 4-layer schedulers (under 3~ms), whereas the delay of \gls{tmrs} is a bit higher but still under 10~ms. For the \gls{udp} fast application \gls{amrs} turns out to be the best in terms of total \gls{dl} throughput, albeit with considerable more delay than \gls{pmrs}. \gls{tmrs} displays high delay and limited throughput due to the lack of resources of the 1-layer frame. The throughput with \gls{pmrs} is about half as much as with \gls{amrs}, but with much lower delay. Notably, the throughput-delay behavior of \gls{pmrs} with \gls{udp} fast is similar to that of the \gls{tcp} application with either 4-layer scheduler. As the \gls{tcp} rate adaptation reduces the transmission window when certain timers expire, its delay is under 30~ms for all schedulers, but the achieved rate with such delay varies. \gls{pmrs} offers the best \gls{tcp} throughput with under 15~ms delay, followed closely by \gls{amrs} scheduler. Finally \gls{tmrs} achieves the worst \gls{tcp} throughput, with the highest delay, due to the limited resources of the 1-layer frame.

The main conclusion of this section is that the \gls{mumimo} system performance depends significantly on the offered traffic. For a lightly loaded cell with fixed traffic, all the configurations discussed offer a satisfactory behavior, whereas strong trade-offs between delay and throughput emerge in an over-loaded cell with fixed traffic. Moreover the different scheduling algorithms diverge significantly in their response to the over-loaded scenario, with \gls{pmrs} displaying better delay generally, \gls{amrs} displaying more \gls{dl} rate with some delay increase, and \gls{tmrs} being overwhelmed by the traffic. Applications on top of \gls{tcp} are more sophisticated and adapt their rates to the network. In this case the severe delays of the over-loaded scenario are avoided by the rate adjustment, which converges to a significantly larger rate for the 4-layer models compared to the 1-layer baseline. Generally \gls{pmrs} offers consistently good performance in both  \gls{ul} and \gls{dl}, whereas \gls{amrs} is a great scheme in \gls{dl} but has severe shortcomings in \gls{ul}.

\section{Conclusions}
\label{sec:conclusions}

In this paper we have studied the simulation of \gls{mumimo} \gls{hbf} implementations for \gls{3gpp} \gls{nr} \gls{mmwave} cellular systems. We have shown that by supporting multiple transmission layers simultaneously, the system capacity is greatly increased. Moreover, by associating each frequency-flat \gls{bf} vector to a separate antenna port, the signal processing involving large arrays characteristic of \gls{mmwave} systems can be handled in a space of reduced dimensions. In addition, by considering a linear matrix mapping logical transmission layers to physical antenna ports, it is possible to leverage the advantages of \gls{mumimo} signal processing techniques in order to alleviate the inter-user interference and improve the \gls{sinr}. We have shown that this is indeed necessary, as the \gls{sinr} would degrade significantly if we merely used separate analog beams for each user without \gls{mumimo}-aware \gls{hbf}. With regard to control overhead, we present a frequency-flat \gls{mmse} \gls{bf} scheme with reduced feedback that achieves a partial interference removal, and a frequency-selective \gls{mmse} \gls{bf} scheme with significantly more feedback that achieves almost complete interference removal.

We have revealed a trade-off between the design of \gls{mumimo} schedulers and the \gls{bf} problem. Particularly, due to the characteristics of channel estimation in \gls{nr}, only coexisting allocations that start at the same time are able to employ \gls{mumimo}-aware \gls{hbf} techniques in order to reduce the interference. This raises a conflict between interference mitigation and \gls{rb} allocation, as some wasteful padding symbols are needed to enforce the constraint that all allocations start at the same time. We have implemented two types of schedulers, one with padding and one that permits asynchronous transmissions and wastes no resources. We have shown through simulation that the latter scheduler leads to system performance degradation on average, although during its operation the events with too much interference are only occasional and may be compensated with adequate retransmission schemes, at the expense of some delay increase.

We have studied the relation between the system throughput and delay performance indicators, the application data rates, and this scheduler-\gls{bf} trade-off. In general the use of the padding scheduler displayed the most consistent behavior, achieving satisfactory delays with much higher throughput than a baseline 1-layer system in both \gls{dl} and \gls{ul}. On the other hand, the asynchronous scheduling approach cannot yet be fully discarded, as we have shown that it offers even greater throughput in some very specific \gls{dl} scenarios. In \gls{ul}, due to the severe outages caused by large differences of pathloss between users, the asynchronous scheduling approach performs poorly.

\vspace{-0.3cm}
\bibliographystyle{IEEEtran}
\bibliography{bibl.bib}

\end{document}